\documentclass[prx,
 reprint,
groupedaddress,
 aps,
floatfix,
]{revtex4-2}
\usepackage{graphicx}
\usepackage{bm}
\usepackage{braket}
\usepackage{amsmath}
\usepackage{amssymb}
\usepackage{tikz}
\usepackage{etoolbox}

\definecolor{myblue}{rgb}{0.12156862745098039, 0.4666666666666667, 0.7058823529411765}
\definecolor{myorange}{rgb}{1.0, 0.4980392156862745, 0.054901960784313725}

\newrobustcmd*{\mycircle}[1]{\tikz{\filldraw[draw=#1,fill=#1] (0,0) circle [radius=0.1cm];}}
\newrobustcmd*{\mytriangle}[1]{\tikz{\filldraw[draw=#1,fill=#1] (0,0) --(0.2cm,0) -- (0.1cm,0.2cm);}}

\usepackage[utf8]{inputenc}
\usepackage{upgreek}
\usepackage[T1]{fontenc}
\usepackage[separate-uncertainty=true,
exponent-product = \cdot,
per-mode = symbol]{siunitx} 
\usepackage{gensymb}
\usepackage{hyperref}
\hypersetup{
    colorlinks=true,
    linkcolor=myblue,
    filecolor=magenta,      
    urlcolor=cyan,
    pdftitle={A compact ion-trap quantum computing demonstrator},
    citecolor = myblue
}

\newcommand{\ca}{$^{40}$\textrm{Ca}\ }
\newcommand{\cai}{$^{40}$\textrm{Ca}$^+$\ }
\newcommand{\MS}{\text{M\o{}lmer-S\o{}rensen}}

\DeclareSIUnit{\dB}{dB}
\DeclareSIUnit{\belmilliwatt}{Bm}
\DeclareSIUnit{\dBm}{\deci\belmilliwatt}
\DeclareSIUnit{\belcarrier}{Bc}
\DeclareSIUnit{\dBc}{\deci\belcarrier}
\DeclareSIUnit{\beli}{Bi}
\DeclareSIUnit{\dBi}{\deci\beli}

\newcommand\Rx[1]{\def\arg{#1}\ifx\arg\empty \ensuremath{\hat{R}_x}\else \ensuremath{\hat{R}_x(#1)}\fi}
\newcommand\Ry[1]{\def\arg{#1}\ifx\arg\empty \ensuremath{\hat{R}_y}\else \ensuremath{\hat{R}_y(#1)}\fi}
\newcommand\Rz[1]{\def\arg{#1}\ifx\arg\empty \ensuremath{\hat{R}_z}\else \ensuremath{\hat{R}_z(#1)}\fi}

\begin{document}

\title{A compact ion-trap quantum computing demonstrator}
\author{I. Pogorelov$^1$}
\author{T. Feldker$^1$}
\author{Ch. D. Marciniak$^1$}
\author{L. Postler$^1$}
\author{G. Jacob$^2$}
\author{O. Krieglsteiner$^2$}
\author{V. Podlesnic$^1$}
\author{M. Meth$^1$}
\author{V. Negnevitsky$^3$}
\author{M. Stadler$^3$}
\author{B. H{\"o}fer$^4$}
\author{C. W{\"a}chter$^4$}
\author{K. Lakhmanskiy$^{1,5}$}
\author{R. Blatt$^{1,6}$}
\author{P. Schindler$^1$}
\author{T. Monz$^{1,2}$}
\email{thomas.monz@uibk.ac.at}

\address{$^1$ Institut f{\"u}r Experimentalphysik, 6020 Innsbruck, Austria}
\address{$^2$ Alpine Quantum Technologies (AQT), 6020 Innsbruck, Austria}
\address{$^3$ Institute for Quantum Electronics, ETH Z{\"u}rich, 8093 Z{\"u}rich, Switzerland}
\address{$^4$ Fraunhofer-Institut für Angewandte Optik und Feinmechanik IOF, 07745 Jena, Germany}
\address{$^5$ Russian Quantum Center, 121205 Moscow, Russia}
\address{$^6$ Institute for Quantum Optics and Quantum Information, 6020 Innsbruck, Austria}

\begin{abstract}
Quantum information processing is steadily progressing from a purely academic discipline towards applications throughout science and industry. Transitioning from lab-based, proof-of-concept experiments to robust, integrated realizations of quantum information processing hardware is an important step in this process. However, the nature of traditional laboratory setups does not offer itself readily to scaling up system sizes or allow for applications outside of laboratory-grade environments. This transition requires overcoming challenges in engineering and integration without sacrificing the state-of-the-art performance of laboratory implementations. Here, we present a 19-inch rack quantum computing demonstrator based on \cai optical qubits in a linear Paul trap to address many of these challenges. We outline the mechanical, optical, and electrical subsystems. Further, we describe the automation and remote access components of the quantum computing stack. We conclude by describing characterization measurements relevant to quantum computing including site-resolved single-qubit interactions, and entangling operations mediated by the {\MS} interaction delivered via two distinct addressing approaches. Using this setup we produce maximally-entangled Greenberger–Horne–Zeilinger states with up to 24 ions without the use of post-selection or error mitigation techniques; on par with well-established conventional laboratory setups.

\end{abstract}

\maketitle
\date{\today}

\section{Introduction}

Quantum information processing as a computational paradigm has been proposed as an efficient means to tackle computational challenges throughout science and industry~\cite{
Deutsch:1992, Shor:1994, Yung:2014,preskill2018quantum, McArdle:2020, Bauer:2020}. The rapid scaling of the computational potential with the number of quantum bits (qubits) is the basis for widespread interest in realizing a quantum information processor, or quantum computer. Tremendous progress has been made both experimentally and theoretically in exploring and demonstrating hallmark capabilities of quantum information processors in numerous architectures~\cite{Arute:2019, Zhang:2017, Bernien:2017, Wang:2018, yang2020operation,pla2012single}. Among the most successful of these architectures are trapped atomic ions, which have already met many requirements for fault-tolerant~\cite{Shor:1996} quantum computation~\cite{Benhelm:2008, Nigg:2014, Ballance:2016, Gaebler:2016,zhang2020error}.

With progressing capabilities among all platforms, the attention has recently shifted away from proof-of-principle implementations towards integration and scalability of architectures~\cite{Benhelm:2008, Monz:2016, Zhang:2017, Figgatt:2019, Arute:2019, Hempel:2018, mehta2020integrated, niffenegger2020integrated}. The shift in development of quantum computers from small-scale, expert-user devices to integrated, end user-centric systems is analogous to the history of classical computation. It presents a host of new challenges such as manufacturing many, effectively identical qubits, and improving the scaling in number of control and readout lines, while maintaining low error rates~\cite{crippa2019gate,jarratt2019dispersive,reilly2019challenges}. The minimization of resource overhead incurred in quantum control techniques or quantum error correction is an ongoing challenge across architectures~\cite{Benhelm:2008}. These challenges are prominent examples of problems that need to be overcome as quantum devices scale to hundreds of qubits. It has become clear over the last decade that any device that can outperform classical high-performance computing for tasks relevant in industrial or applied science settings will require substantially more qubits than current architectures can sustain~\cite{clark2009resource, wecker2014gate,takeshita2020increasing,babbush2019quantum,moll2016optimizing}. In addition to a large number of qubits, such a device should present the user with a hardware-agnostic interface, and require little to no maintenance by on-site personnel during standard operation. Meanwhile, all the basic routines should be automated and perform at the level sufficient for fault-tolerance requirements. Finally, the device should be deployable outside of well-controlled, low-noise laboratory conditions with purpose-built infrastructure. Demonstrating the capabilities of scalable architectures beyond laboratory implementations is therefore a crucial next step.

In this work we present the first phase of our efforts towards a compact, 50-qubit quantum computing demonstrator based on trapped ions as part of the AQTION (Advanced Quantum computation with Trapped IONs) collaboration. It features a $\SI{1.7}{\meter}\times\SI{1}{\meter}$ footprint with high mechanical stability, and scalable control electronics. We describe the hardware concept and integration, and characterize the initial system performance including entangling gate operation necessary for quantum computation.

For a quantum computer to perform arbitrary computations it is sufficient to implement arbitrary-pair, two-qubit entangling gates in addition to single-qubit rotations~\cite{brylinski2002mathematics, bremner2002practical}. In a trapped ion quantum computer these operations are typically the result of interactions with laser fields. Addressing individual qubits in a register with light beams is thus an essential component of universal quantum computation efforts in trapped ions~\cite{naegerl1999laser, warring2013individual}. Meeting the demands of state preparation and measurement~\cite{mavadia2013control, goodwin2016resolved, Stricker:2017} in a scalable fashion is therefore a major challenge in trapped ion quantum computing. Consequently, the demonstrator leverages industrial expertise to accomplish scalable integration of light generation and delivery, in particular single-site addressing essential for the trapped-ion implementation. The demands of quantum control and algorithmic compiling on the software stack with increasing qubit and gate numbers are similarly challenging~\cite{chong2017programming}. A detailed description of the holistic software stack of the AQTION platform, spanning device-specific control to hardware-agnostic end user algorithms, is beyond the scope of the present paper and will be covered in upcoming publications.

This manuscript is structured as follows: In section \ref{sec:PhysicalImplementation} we present an overview of the physical qubit implementation, as well as the means of control, preparation, readout, and manipulation as necessary for computation. In section \ref{sec:TechnicalImplementation} we describe the main subsystems by functional groups, including mechanical, optical, and electrical subsystems, as well as automation features of the demonstrator. In section \ref{sec:Characterization} we turn to characterization measurements on the composite system. This manuscript concludes in section \ref{sec:Conclusion}, where we outline near-term hardware integration goals to expand this setup into a fully self-contained, trapped ion-based quantum computing demonstrator.
	
\section{The qubit system}
\label{sec:PhysicalImplementation}
The choice of atomic species in a trapped-ion experiment is always based on trade-offs between beneficial physical properties of a given species, and technical implementation challenges associated with it. Broadly speaking, atomic qubits can be either optical qubits, Zeeman qubits or hyperfine qubits. In optical qubits quantum information is encoded in two electronic states connected by an electric multipole transition with frequency in the optical domain, and the excited state is long-lived or meta-stable. In Zeeman or hyperfine qubits the information is encoded in magnetic sublevels of electronic ground states with transition frequencies in the microwave to radiowave domain. A species may host several different types of qubits  distinct in their implementation. Each species and qubit type offers advantages and disadvantages which may benefit certain applications more than others. At this stage no single species or qubit type has been identified as ultimately superior. The design goals and principles for our trapped-ion demonstrator are largely independent of the choice of ion species or qubit type to reflect the flexibility that this fact requires.

In our particular demonstrator we use optical qubits encoded in the electronic state of the valence electron in nuclear spin-free \cai ions. This choice is motivated in part by technical considerations such as commercially available semiconductor laser sources and high-quality optics for all the required transitions. Transitions wavelengths are in the blue, red and infrared part of the optical spectrum. Compared to transitions wavelength in the ultraviolet (UV) this has several advantages such as reduced charging of trap electrodes and substantially lower onset of solarization or photodarkening of optical components. Optical qubits can be directly interacted with using only a single beam compared to more complex beam geometries for Raman gates in Zeeman qubits or hyperfine qubits.

\begin{figure}[h!]
    \centering
    \includegraphics[width=8.6cm]{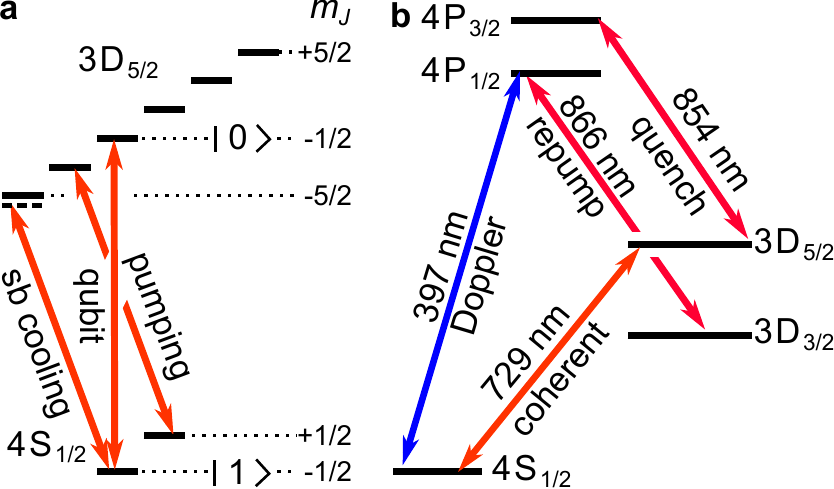}
    \caption[Concept]{Energy levels and relevant transitions in \cai for quantum computing demonstrator. \textbf{a} Ground and excited state fine structure manifold with electric quadrupole transitions for sideband (sb) cooling on the first red sideband, optical pumping, and the $\Delta m_J=0$ qubit transition. \textbf{b} Ground, excited and auxiliary levels in \cai with transitions for Doppler cooling and detection, repumping, and quenching of the excited state lifetime during sideband cooling.}
    \label{fig:fig_LevelScheme}
    \end{figure}

Specifically, the qubit $\ket{1}$ state is the $\ket{\textrm{4\,S}_{1/2},m_J = -1/2}$ Zeeman state, which is coupled to the long-lived qubit $\ket{0}$ state $\ket{\textrm{3\,D}_{5/2},m_J = -1/2}$. This excited state has a lifetime of $\tau = \SI{1.168(7)}{\second}$~\cite{Barton:2000}, and decays via an electric quadrupole transition near \SI{729}{\nano\meter}, see Fig.~\ref{fig:fig_LevelScheme} \textbf{a}. This transition has the lowest magnetic field sensitivity in the manifold (\SI{5.6}{\mega\hertz\per\milli\tesla}) and is suitable for an effective two-level system as shown in Figs~\ref{fig:fig_LevelScheme} \textbf{a} and \textbf{b}.

The high magnetic field sensitivity compared to clock transitions in hyperfine qubits can be mitigated by proper magnetic shielding and stabilization~\cite{ruster2016long}. 
A more fundamental limit of this optical qubit used is the lifetime of the upper qubit state $\ket{\textrm{3\,D}_{5/2},m_J = -1/2}$. However, $\tau$ is about 4 orders of magnitude longer than typical 2-qubit operations (\SI{200}{\micro \second}) and 5 orders of magnitude longer than single-qubit operations (\SI{15}{\micro \second}). Thus, in the medium term, gate fidelity is not limited by the fundamental physical properties of the qubit but by the specifics of the technical implementation.

Preparation of the qubit system in our demonstrator entails two tasks: First, loading of ions into the trap is typically performed only when the qubit count has to be changed. Second,  preparation of the electronic and motional state of the ions which is performed before every experiment.

\paragraph{Ion loading} \cai is loaded into the trap either from a thermal source or via ablation from a target. Either method produces ions with temperatures ranging from hundreds to thousands of Kelvin. Atomic \ca features a low-lying excited state connected to the ground state via a dipole transition, which enables isotope-selective, resonantly-enhanced two-photon photoionisation at commercially available laser wavelengths near \SI{423}{\nano\meter} and \SI{375}{\nano\meter}. The ionization process takes place at the center of the trap, where ions are confined afterward by the trapping potential. The ions are then Doppler cooled via an electric dipole cycling transition from 4\,S$_{1/2} \leftrightarrow \textrm{4\,P}_{1/2}$ at \SI{397}{\nano\meter}. A repumping laser near \SI{866}{\nano\meter} prevents population trapping in the metastable 3\,D$_{3/2}$ manifold. Doppler cooling alone may be sufficient to reach crystallization into a Coulomb crystal depending on ion number and confining potential strength. Such Coulomb crystals of $N$ ions support $3N$ bosonic motional modes, $N$ of which are parallel to the weakly-confining axial trap direction, and $2N$ perpendicular to that in the radial trap directions~\cite{thompson2015ion}.

\paragraph{State preparation and readout} State preparation proceeds in four steps. First, ions are Doppler cooled at \SI{397}{\nano\meter} and repumped near \SI{866}{\nano\meter}.  The motional modes of the ion cloud or Coulomb crystal after Doppler cooling will in general be in or close to the Lamb-Dicke regime~\cite{Leibfried_2003}. Second, polarization gradient cooling (PGC) is employed using two counter-propagating, orthogonally-polarized beams blue-detuned from the 4\,S$_{1/2} \leftrightarrow \textrm{4\,P}_{1/2}$ transition~\cite{Ejtemaee:2017}. Polarization gradient cooling results in a thermal state of all motional modes of typically a few quanta of motion for common trapping parameters much below the Doppler cooling limit \cite{Castin1989}. Optical pumping of the qubit manifold on the $\ket{\textrm{3\,D}_{5/2},m_J = -3/2} \rightarrow \ket{\textrm{4\,S}_{1/2},m_J = +1/2}$ transition redistributes the electronic population from the initial mixed population of Zeeman sublevels to the $\ket{\textrm{4\,S}_{1/2},m_J = -1/2}$ state. The final step of preparation is sideband cooling on selected motional modes close to the ground state. The targeted modes in sideband cooling may consist of subsets of the $N$ axial modes, or of the $2N$ radial modes closest to the selected carrier transition that implements a gate operation. The cooling rate in sideband cooling is increased by quenching the lifetime of the 3\,D$_{5/2}$ manifold through coupling to the short-lived  $\textrm{4\,P}_{3/2}$ level via excitation at \SI{854}{\nano\meter}. State-selective readout is performed optically using fluorescence measurements on the Doppler cooling transition, either site-resolved using a camera, or collectively (without spatial resolution) using an avalanche photodiode (APD).

\paragraph{State manipulation} The universal gate set employed~\cite{Maslov_2017} in the demonstrator is spanned by arbitrary single-qubit operations and the two-qubit entangling gate provided via the bichromatic {\MS} gate~\cite{Sorensen:1999, Benhelm:2008}. Both single- and two-qubit gates are implemented using laser light fields focused onto the ions. The effective Hamiltonian governing (near) resonant single-qubit interactions in the demonstrator is given by~\cite{haffner2008quantum} 
\begin{equation*}
H = \hbar \Omega \sigma_{+} \text{e}^{-i\left((\omega - \omega_{eg}) t - \varphi\right)} \text{e}^{i\eta(a \text{e}^{-i\nu t}+a^\dagger \text{e}^{i\nu t})} +h.c.,    
\end{equation*}

where $\omega$ and $\varphi$ denote laser field frequency and phase, $\nu$ is the motional mode frequency, $\omega_{eg}$ is the qubit transition frequency, and $h.c.$ denotes the hermitian conjugates. Further, $\Omega$ denotes the Rabi frequency associated with the transition, $\eta$ is the Lamb-Dicke parameter, $a^\dagger$ is the phonon creation operator, and $\sigma_+$ denotes the atomic (spin) raising operator. This Hamiltonian assumes a well-isolated two-level system for the interaction, uses the rotating wave approximation \cite{Leibfried_2003}, and we neglect all other vibrational modes. This Hamiltonian in first-order Lamb-Dicke approximation leads to a single-qubit unitary propagator of the form
\begin{equation*}
    R(\alpha) = \text{e}^{-i\frac{\alpha}{2}\vec{n}\cdot\vec{\sigma}},
\end{equation*}
where $\alpha$ is an angle that depends on the interaction strength and time, $\vec{n}$ is a unit vector, $\vec{\sigma} = \left\{\sigma_x, \sigma_y, \sigma_z\right\}$ is the Pauli spin vector, and $\vec{n}\cdot\vec{\sigma}$ quantifies the strength of the interaction along the three spin directions. In the atomic eigenbasis this can be expressed in spherical-coordinate form assuming $\omega=\omega_{eg}$.
\begin{equation*}
R(\theta, \phi) = 
\begin{pmatrix}
    \cos \theta/2 & -i \text{e}^{-i\phi} \sin \theta /2 \\
    -i \text{e}^{i\phi} \sin \theta /2 &  \cos \theta/2
\end{pmatrix},
\end{equation*}
where $\theta$ and $\phi$ can be controlled via changing the amplitude or interaction time of the laser field, and its phase. The propagator lends itself to the interpretation of rotating the state of a qubit on the Bloch sphere, and single-qubit operations are consequently referred to as single-qubit rotations.

Qubit-qubit interactions in linear ion chains forming along the weakly confining trap axis are mediated via the bosonic bus of shared motional modes through spin-dependent optical dipole forces~\cite{Cirac:1995}. These forces are generated via a bichromatic laser field detuned slightly from the upper and lower sideband transition of a selected vibrational mode. The effective Hamiltonian governing this interaction is given by
\begin{align*}
    H = \sum_{j=j_1,j_2} \sum_{k=1,2}& \hbar \Omega_{j k} \sigma_{j+} \text{e}^{-i((\omega_k - \omega_{eg})t - \varphi_k)}\\
    & \text{e}^{i\eta_j (a \text{e}^{-i\nu t}+a^\dagger \text{e}^{i\nu t})} + h.c.,
\end{align*}
where $j$ enumerates ions, and $k$ enumerates the laser tones. Then, $\omega_{k}$ and $\varphi_k$ denote frequencies and phases of the bichromatic laser field, $\nu$ the closest motional mode frequency, $\eta_j$ the Lamb-Dicke parameter for this mode, $\omega_{eg}$ the frequency of the qubit transition, and $\Omega_{jk}$ the Rabi frequencies of the $k$th beam for the $j$th ion. As before, $a^\dagger$ is the phonon creation operator, $\sigma_{j+}$ is the atomic (spin) raising operator. In the Lamb-Dicke approximation this leads to a two-qubit unitary propagator of the form
\begin{equation*}
    U_\text{MS}(\chi) \approx \text{e}^{-i\chi S_x^2} =
    \begin{pmatrix}
    \cos \chi & 0 & 0 & -i \sin \chi \\
    0 & \cos \chi & -i \sin \chi & 0 \\
    0 & -i \sin \chi & \cos \chi & 0 \\
    -i \sin \chi & 0 & 0 & \cos \chi
\end{pmatrix},
\end{equation*}
where $\chi$ can be controlled with laser field power or interaction time and $S_x = \sum_j \sigma_{j,x}$ is the total spin along $x$. The bichromatic beam parameters should obey certain relations to guarantee that the motional state is disentangled from the ions' spin states at the end of the interaction~\cite{sorensen2000entanglement, roos2008ion}. 

\section{Technical implementation}
\label{sec:TechnicalImplementation}

Demonstrations of many of the requisite capabilities for quantum computation using trapped ions have been presented using laboratory setups~\cite{Benhelm:2008, Nigg:2014, Ballance:2016, Gaebler:2016,zhang2020error}. However, different design constraints apply when constructing traditional laboratory setups or our modular approach. In the context of this work, we highlight the following: (i) The demonstrator needs to minimize footprint while maintaining mechanical stability to move towards scalability without sacrificing performance. (ii) Implementing modularity with flexible standard interconnects helps in reducing footprint and increasing rigidity, while also allowing replaceability and reconfiguration without major redesign, or manual realignment. (iii) The demonstrator needs to be self-contained and rely as little as possible on purpose-built infrastructure. This puts restrictions on external supplies like power, cooling, or environmental conditioning of the demonstrator location. (iv) Scalability inevitably requires standardization. Utilizing established standards to leverage industrial processes and interfaces is therefore desirable. (v) Hardware-agnostic design and scalability require use of automation, as well as remote operability.

\begin{figure*}
		\centering
		\includegraphics[width=13.07cm]{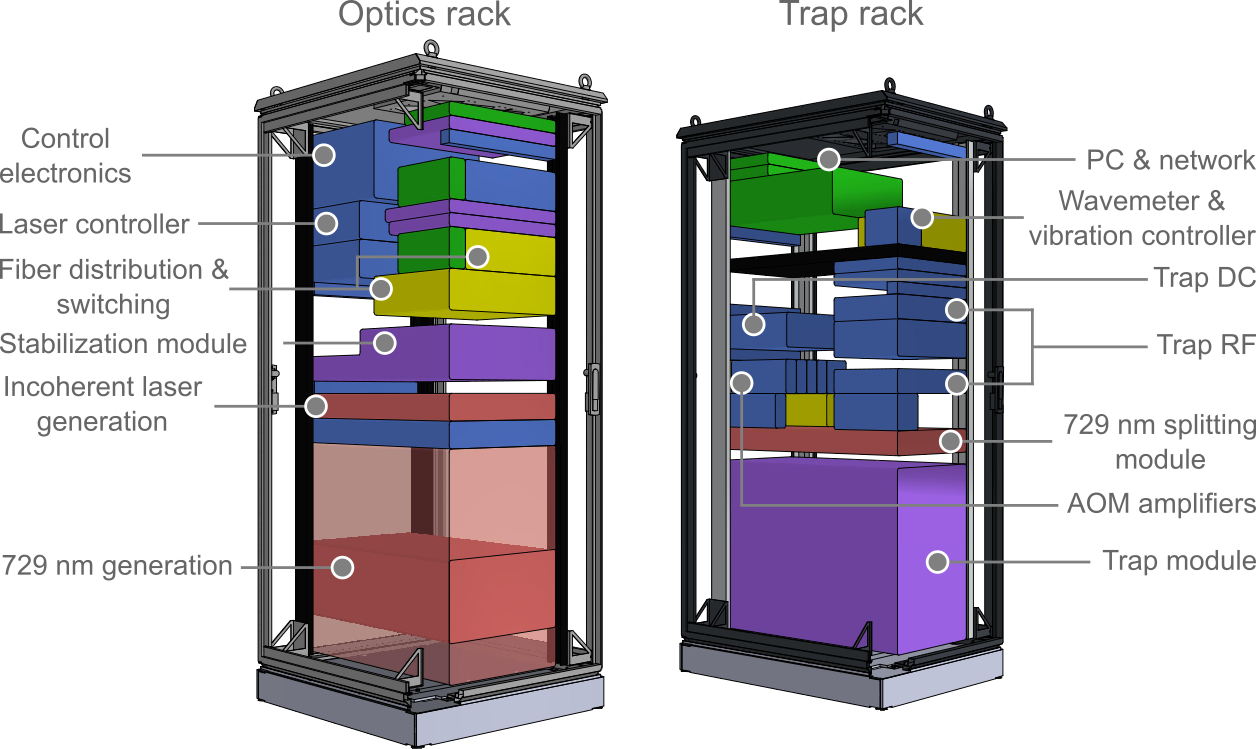}
		\caption[Concept]{Simplified scale model of quantum computing demonstrator housed in two 19-inch racks with major components labelled. Modules in red correspond to optical systems, green for communication and readout, blue electronics and amplifiers, yellow fiber routing and switching, and purple for miscellaneous core modules. The 'optics rack' contains primarily light generation, switching and routing modules with associated electronics. It additionally houses the coherent radio frequency (RF) and digital signal generation module. The 'trap rack' houses the main trap module with associated drive electronics, as well as the communications and remote control hub. Interconnects between modules and racks via electrical and optical patch cords. Semi-transparent red is planned \SI{729}{\nano\meter} light generation module.}
		\label{fig:Racks}
	\end{figure*}

In this section we present our composite demonstrator's setup whose overall concept is shown in Fig.~\ref{fig:Racks}. The demonstrator setup is contained within two industry standard 19-inch racks. Connections between the modules inside each rack and between the two racks are achieved via electrical and optical patch cords, respectively. One of the two racks primarily houses modules related to generation, switching and routing of the required optical frequencies and is hence designated \textit{optics rack}. The second houses the ion trap, associated infrastructure and drive electronics, thus being designated the \textit{trap rack}. 
	
\subsection{Mechanical assembly and environmental conditions}
The primary mechanical assembly is composed of two industry standard 19-inch racks with a footprint of $\SI{0.85}{\meter}\times\SI{1}{\meter}$ each at a height of \SI{2}{\meter}. Modules are largely free of moving parts to improve mechanical rigidity, and long-term stability. Modules that require manual alignment are equipped with self-arresting slide drawers, such that maintenance does not necessitate disassembly. The sole external supply to the racks is one standard \SI{16}{\A}/\SI{230}{\volt} connector per rack, for a total power consumption of less than \SI{3.7}{\kilo\W} per rack. Temperature control inside the racks is provided by forced-air cooling and passive cooling fins throughout the modules. Air flow is restricted across free-space optical parts by appropriate enclosures. This prevents beam pointing instability and phase-fluctuations induced by the moving air column. The racks are closed with doors in standard operation, improving directionality of air flow for better cooling, and protecting the equipment from dust.

\subsubsection{Ion trap and vacuum apparatus}

The \cai ions are confined in a linear Paul trap (AQT Pine trap) consisting of electrodes machined from gold-plated titanium, and an alumina holder which serves as mounting and electrical isolation between the electrodes. The macroscopic trap design of the demonstrator consists of four blade electrodes and two endcaps, and is a variant of earlier designs employed previously in the Innsbruck ion trapping group~\cite{Guggemos:2015, Schindler:2013}. The distance from the center of the trap to the surface of the blade electrodes is $r_0 = \SI{0.57}{\milli\m}$, while the endcap-endcap distance is $z_0 = \SI{4.5}{\milli\m}$.

Radio frequency (RF) voltage is applied to two opposing blade electrodes, and a positive static (DC) voltage to the endcaps. A set of additional electrodes allows for compensation of stray electric fields in order to reduce excess micromotion. The remaining two blade electrodes are held at a DC potential, but may be supplied with an out-of-phase RF potential for symmetric drive, if desired for complete cancellation of axial micromotion. The trap features a thermal calcium source (oven) for photoionization, as well as an ablation target. An \textit{in vacuo} PT100 temperature sensor allows for temperature monitoring of the trap in operation.

The trap assembly uses exclusively non-magnetic materials, specifically titanium, copper, alumina and austenitic stainless steel (grade 1.4429 or equivalent) to minimize distortion of the magnetic environment.
	
The Paul trap itself is located inside a compact stainless steel spherical octagon vacuum chamber. Six diametrically opposing, non-magnetic fused silica DN40 viewports with an antireflective (AR) coating allow for low numerical aperture (NA) optical access on the trap perimeter with NA $\approx 0.05$. Additionally, a viewport with a clear aperture of \SI{44.2}{\milli\m} in the trap mounting flange provides optical access with a medium NA $\approx 0.29$ for imaging of ions to an avalanche photodiode. From the opposing side of the vacuum chamber a re-entrance viewport with clear aperture of \SI{71.2}{\milli\m} at a distance of \SI{18.3}{\milli\m} to the trap center provides optical access via a $\textrm{NA} = 0.6$ objective lens (Photon Gear, 18WD Atom Imager Objective) for resolved ion imaging and illumination.
	
The trap mounting flange is equipped with feedthroughs for electrical connection to the trap electrodes, calcium oven and the PT100 temperature sensor. Two non-evaporative getter (NEG) pumps (SAES NexTorr Z200) provide the main pumping capacity to the chamber. A small ion getter pump (SAES CapaciTorr Z100) additionally pumps noble gases and provides pressure monitoring. Material selection for chamber construction was again restricted to low magnetic permeability, with the exception of permanent magnets required for pump operation.
	
Permanent magnets are further used in Halbach and Helmholtz configuration to provide the quantization field for the qubits. These are mounted directly to the vacuum chamber. The resulting homogeneous field in the trap center has an angle of 66\degree\ and 34\degree\ to the trap axis, and the imaging axis, respectively. The magnitude of the total field produced is $B_0 = \SI{0.50}{\milli\tesla}$. 
Temperature-compensated samarium cobalt magnets are employed to reduce magnetic field fluctuations to a minimum. Their relative change in magnetic field strength with temperature is $\delta_T B \approx \SI{1e-5}{\kelvin^{-1}}$.  Furthermore, three sets of compensation coils along three orthogonal axes allow for control and stabilization of the magnetic field at the ion location in both absolute value and gradient.
	
\subsubsection{Trap support infrastructure and environmental isolation}
	
\begin{figure*}[t]
    \centering
    \includegraphics[width=13cm]{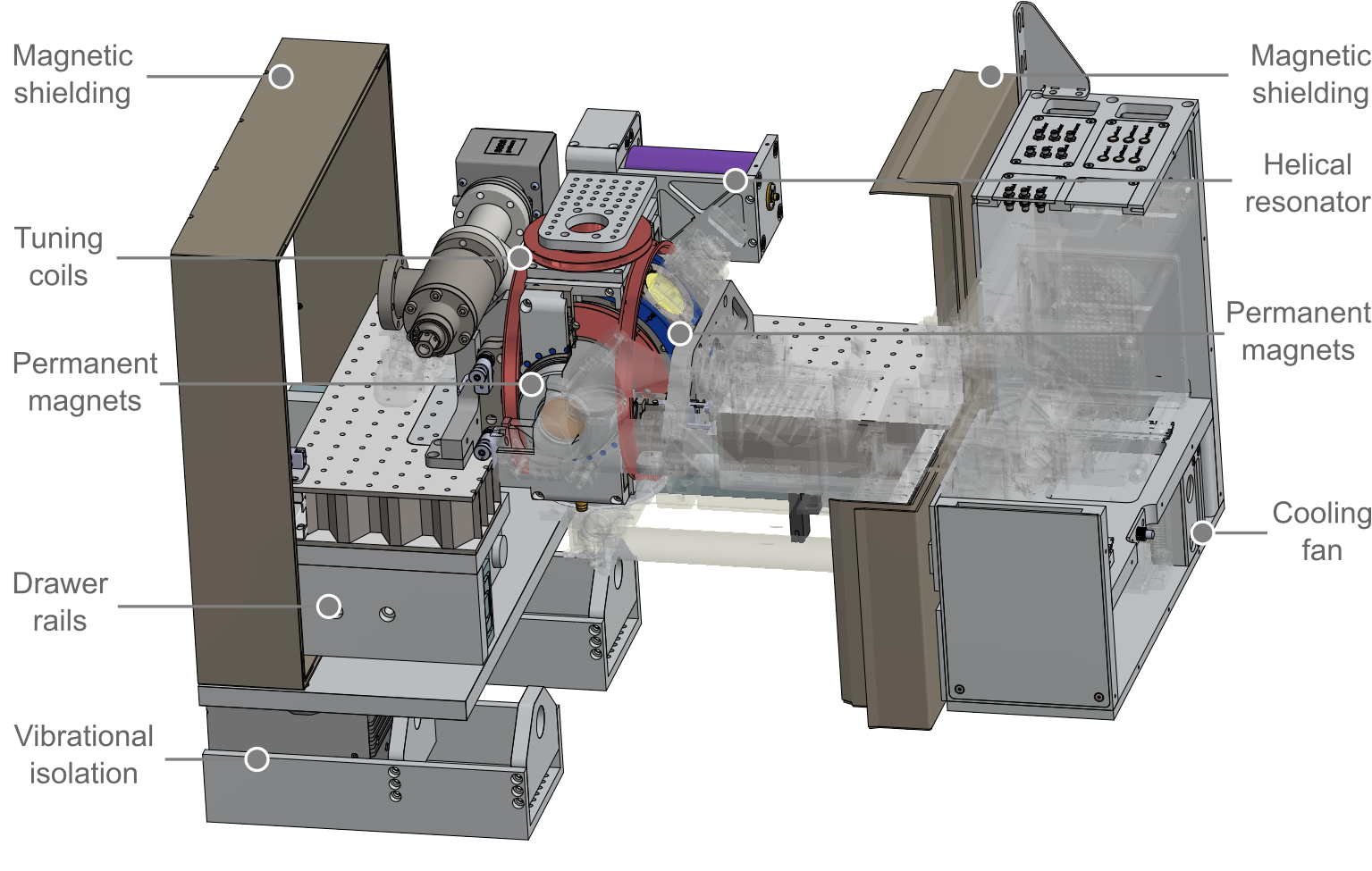}
    \caption{Schematic of trap drawer detail, focusing on the trap proper and mechanical components, displayed with extended (opened) drawer. Three sets of compensation coils with each a Helmholtz and anti-Helmholtz coil per holder (red) together with four sets of permanent magnets in Helmholtz (blue, left leader) and Halbach (blue, right leader) configuration define and compensate the magnetic environment. $\upmu$-metal shielding encloses the trap setup except for penetrations allowing electrical and optical access. Active vibration isolation and thick honeycomb-lattice breadboards provide mechanical stability. A fan outside the shielding provides cooling for components close to the shield.}
    \label{fig:TrapDrawer_Mechanical}
\end{figure*}

The Paul trap and accompanying vacuum chamber are the most critical components with respect to environmental disturbances like thermal, mechanical or magnetic field fluctuations. The employed \cai implementation maintains sensitivity to magnetic field noise that can limit attainable coherence times. Qubit operations depend on amplitude, phase and frequency of the interacting light fields. Therefore, any fluctuations on those quantities, as for example caused by beam pointing instabilities, will adversely affect operation fidelity. Consequently, these critical components are situated in an environmentally isolated housing, as shown in Fig.~\ref{fig:TrapDrawer_Mechanical}, which we refer to as the \textit{trap drawer}.
The drawer itself, designed at AQT, makes up the bottom layer of the trap rack, and is fastened to it via supports. The supports in turn connect to sliding rails on ball bearings through active vibration damping elements (Accurion Vario) based on pre-loaded springs and magnetic repulsion elements. Measurements of the isolators' magnetic fields show that their influence at the ions' position is below ambient magnetic field level. The sliding drawer allows for easy access during installation and maintenance by clearing the main rack structure. The majority of the drawer footprint is enclosed in a $\upmu$-metal (ASTM A753 Alloy 4) shield (supplied by Magnetic Shield Ltd) for magnetic isolation, which has a limited number of penetrations for electrical and optical access. Active instruments and electronics that can cause magnetic field noise are outside the shielded section wherever feasible.

\subsection{Optical subsystems}

Interaction with and readout of the qubits is primarily undertaken using optical fields in the presented demonstrator. The ability to manipulate and read out single qubits with light is thus central to its operation. Optical and optoelectronic subsystems consequently occupy an appreciable portion of the available space and complexity. In the following we will discuss the generation and stabilization of the requisite lasers, present the light routing and addressing capabilities, and finally summarize the readout and detection components.

\subsubsection{Laser light generation and stabilization}

\begin{figure}[h!]
		\centering
		\includegraphics[width=8.6cm]{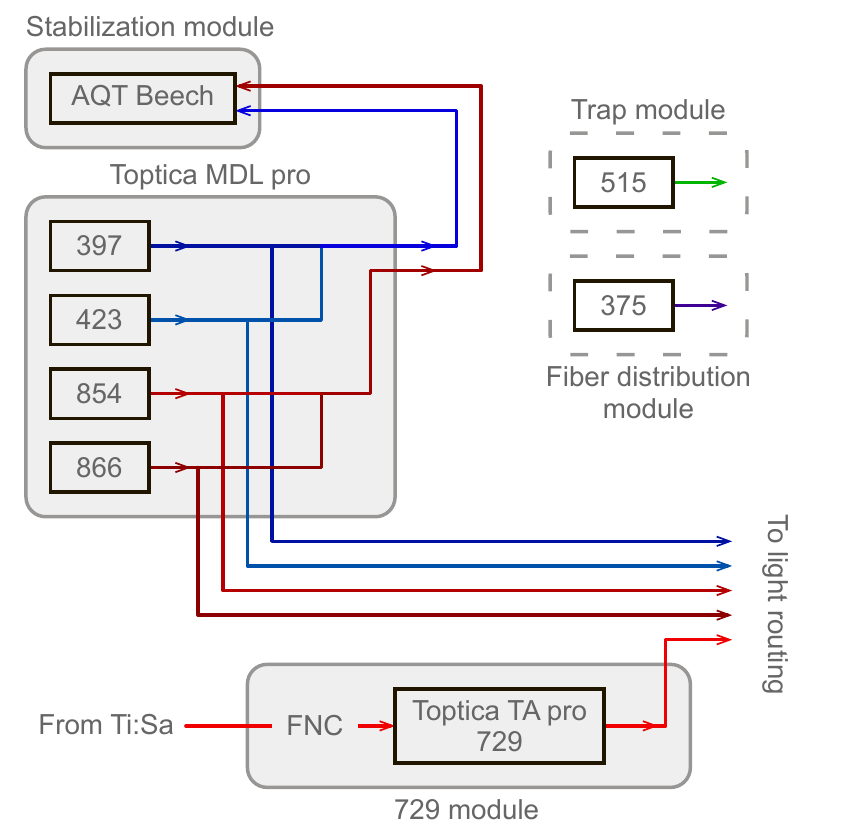}
		\caption[Light generation]{Schematic of principal light generation modules. A modified Toptica MDL pro unit produces four of the five laser colors needed for incoherently-driven excitations. Their output is fiber-coupled to a frequency stabilization module. A free-running laser diode is additionally installed downstream to provide the final incoherent excitation color. Laser colors within the qubit manifold are generated by a Toptica TA pro seeded by light from an ultra-stable MSquared SolsTiS Ti:Sa master oscillator present in the laboratory.}
		\label{fig:LightGeneration}
	\end{figure}

The demonstrator is equipped with three principal laser generation modules, as indicated in Fig.~\ref{fig:LightGeneration}. First, a Q-switched (pulsed) laser (Coherent Flare NX) situated outside the magnetic shielding in the trap drawer emitting at \SI{515}{\nano\meter} provides highly energetic pulses for ablation loading~\cite{sheridan2011all} of \cai from the \textit{in vacuo} target. Second, a multi-color diode laser (MDL) generation module (Toptica, MDL pro) provides light for all incoherently-driven interactions: Doppler and polarization gradient cooling at \SI{397}{\nano\meter}, line broadening ('lifetime quenching' for the qubit reset) at \SI{854}{\nano\meter}, repumping at \SI{866}{\nano\meter}, and \SI{423}{\nano\meter} light used in the first (resonantly enhanced, isotope-selective) step for photoionization. This MDL setup is supplemented by a free-running laser diode (Toptica iBeam smart) providing light at \SI{375}{\nano\meter} for the second (non-resonant) photoionization step to the continuum. The last major light generation module provides light at \SI{729}{\nano\meter} for sideband cooling, optical pumping, and the coherent qubit control transition. Preliminary operation of this module uses a tapered amplifier (Toptica, TA pro), which is fed by seed light generated outside of the rack until the integrated laser source is installed. The seed light is derived from an ultra-stable titanium sapphire (Ti:Sa) master oscillator (MSquared SolsTiS) locked to a high-finesse optical cavity, resulting in a linewidth of \SI{3.6(4)}{\hertz}~\cite{Stricker:2017}. The seed light is fed into the TA via optical fiber with active fiber noise cancellation~\cite{Ma:1994}. The TA has an output power of \SI{500}{\milli\watt} at a seed level of \SIrange{15}{30}{\milli\watt}, which is sufficient to drive the amplifier gain into saturation. The bichromatic light fields required for the {\MS} interaction are generated by supplying a two-tone RF signal to the acousto-optic modulator (AOM) further downstream.
Future hardware upgrades will be composed of a compact, diode-based laser system stabilized to a high-finesse cavity, which is completely contained in the rack. The pending system upgrade features a specified linewidth of \SI{<2}{\hertz} and an output power of \SI{>200}{\milli\watt} after fiber noise cancellation. 

Both ablation loading and photoionization inherently have no need for stringent frequency stabilization. On the other hand, all interactions with \cai require precise control of the absolute frequency of the respective lasers to maintain efficient operation. The MDL unit has therefore been extended to provide two fiber-coupled feedback outputs combining all four frequencies on two patch cords, which are fed to a dedicated stabilization module. The stabilization unit should provide reliable low-noise frequency locking along with low long-term drift rates. In addition to these general requirements we need to comply with the demonstrator's design principles, which demand compactness, remote control access, and automation features. The module used here to comply with these requirements is the AQT Beech module. Inside, all lasers are locked to a reference cavity. The cavity in turn is stabilized via a transfer lock to an atomic transition in a gas cell. The details of this system will be presented in a separate publication.
	
\subsubsection{Light delivery, switching, and addressing}
\label{subsec:LightDelivery}

\begin{figure}[h!]
		\centering
		\includegraphics[width=8.6cm]{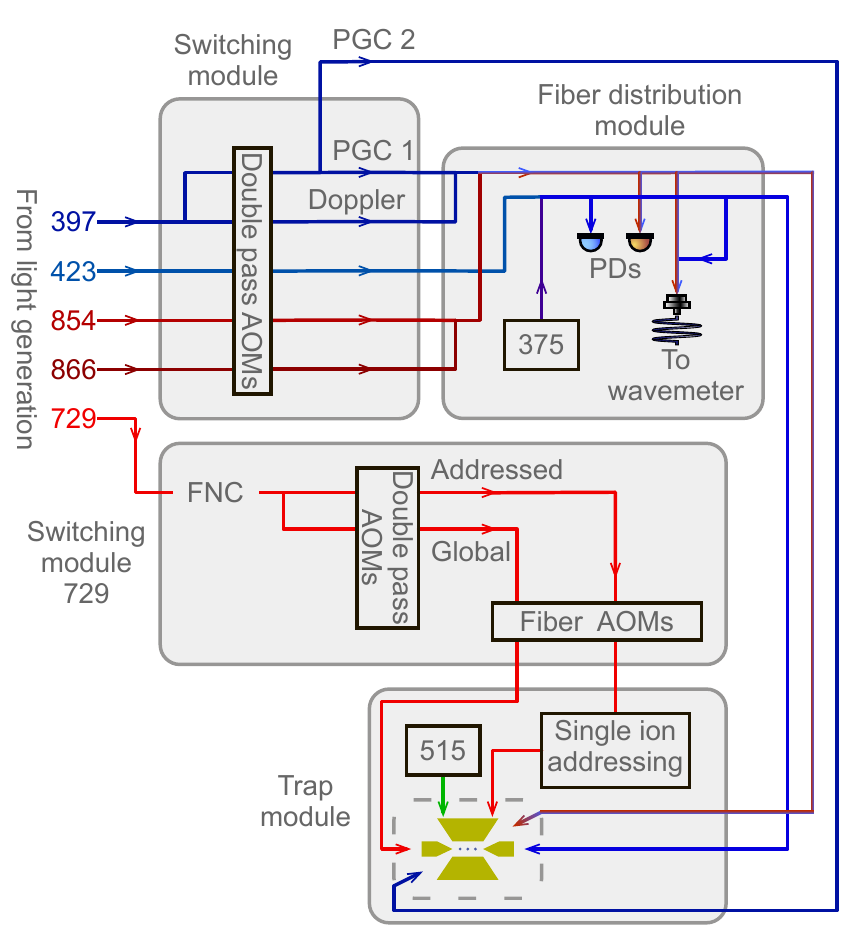}
		\caption[Delivery]{Schematic of light delivery, switching and addressing modules. Light is fiber-delivered to dedicated switching modules utilizing free-space double-pass AOMs, after which overlapping and fiber routing to their final destination is handled on a fiber distribution module for incoherent interactions. Photodiodes (PDs) are placed throughout the beamline for continuous power monitoring. The qubit laser system has similar switching and routing capabilities, with addition of fiber noise cancellation (FNC), and fiber AOMs for improved on/off extinction ratio, bichromatic modulation for {\MS} interactions, and light field parameter control. One of two approaches is used to deliver addressed light, see Figs.~\ref{fig:BeamLayout} and~\ref{fig:Addressing_Overview}.}
		\label{fig:LightDelivery}
	\end{figure}

The delivery subsystem handles the necessary routing and spatial distribution of the light colors throughout the setup. This includes collective delivery for all the lasers as well as site-selective (addressing) delivery for coherent qubit control. The system provides control over the temporal and intensity profile of the light fields via amplitude shaping, amplitude stabilization and amplitude switching. The optical subsystems have been made modular whenever possible to provide replaceability and readjustment without disturbing other parts of the setup. Inter-module connection is achieved via fiber-optical patch cords in line with the requirements on stability and modularity. Polarization-maintaining single-mode optical fibers are used for delivery throughout due to the polarization sensitivity of optical transitions between Zeeman sublevels in the magnetic field. Typically, free-space optics outperform fiber optics in terms of transmission loss and polarization extinction ratio. This gap in performance grows for shorter wavelengths. Consequently, intermediate free-space optics are employed where fiber-based components perform insufficiently. Figure~\ref{fig:LightDelivery} shows a schematic of the delivery subsystem.

Four main delivery points for optical access are arranged orthogonal to the high NA access ports of the vacuum chamber. Each of these is equipped with a fiber collimator for delivery. The collective qubit laser at \SI{729}{\nano\meter} following the FAOM propagates through holes in the trap endcaps. On the opposing side is another single-mode fiber with the two photoionization colors. Delivery of all other collective beams happens at 45\degree\ to the trap axis. A large-mode-area photonic crystal fiber (PCF) (NKT Photonics LMA-PM-5) delivers the superimposed Doppler cooling laser, refreeze laser, one of the two PGC lasers at \SI{397}{\nano\meter}, as well as the repumping and quench lasers at \SI{866}{\nano\meter} and \SI{854}{\nano\meter}, respectively. A single-mode fiber on the opposing port delivers the remaining second PGC laser to complete the scheme. The high energy \SI{515}{\nano\meter} light pulses for ablation loading are coupled free-space into the trap from below. The four fiber access ports are equipped each with a fast and slow photodiode for monitoring power and stabilization close to the ion position.
	
Individual control over the amplitude of not only each color, but each laser beam, is required by nature of the gate-based interaction. Amplitude shaping is implemented via free-space double-pass AOMs. These shaping AOMs further provide the individual frequency offsets required to bridge the detuning between frequency stabilization cavity resonance and required transition frequency. They are situated inside dedicated rack units as part of the switching module. An additional mechanical shutter is inserted in the Doppler cooling beam path, normally blocking the undiffracted 0th order after the switching AOM. Opening the shutter gives access to a laser which is red-detuned by $\approx\SI{300}{\mega\hertz}$ from the main cycling transition, and can be used to re-freeze a molten ion crystal. The light switching modules are followed by a dedicated fiber distribution module, where free-space optics are used to overlap and fiber-couple beams into patch cords for final delivery. The fiber distribution and switching modules further incorporate photodiodes for continuous power monitoring on all colors throughout the beamline.

\begin{figure}
    \centering
    \includegraphics[width = 8.6cm]{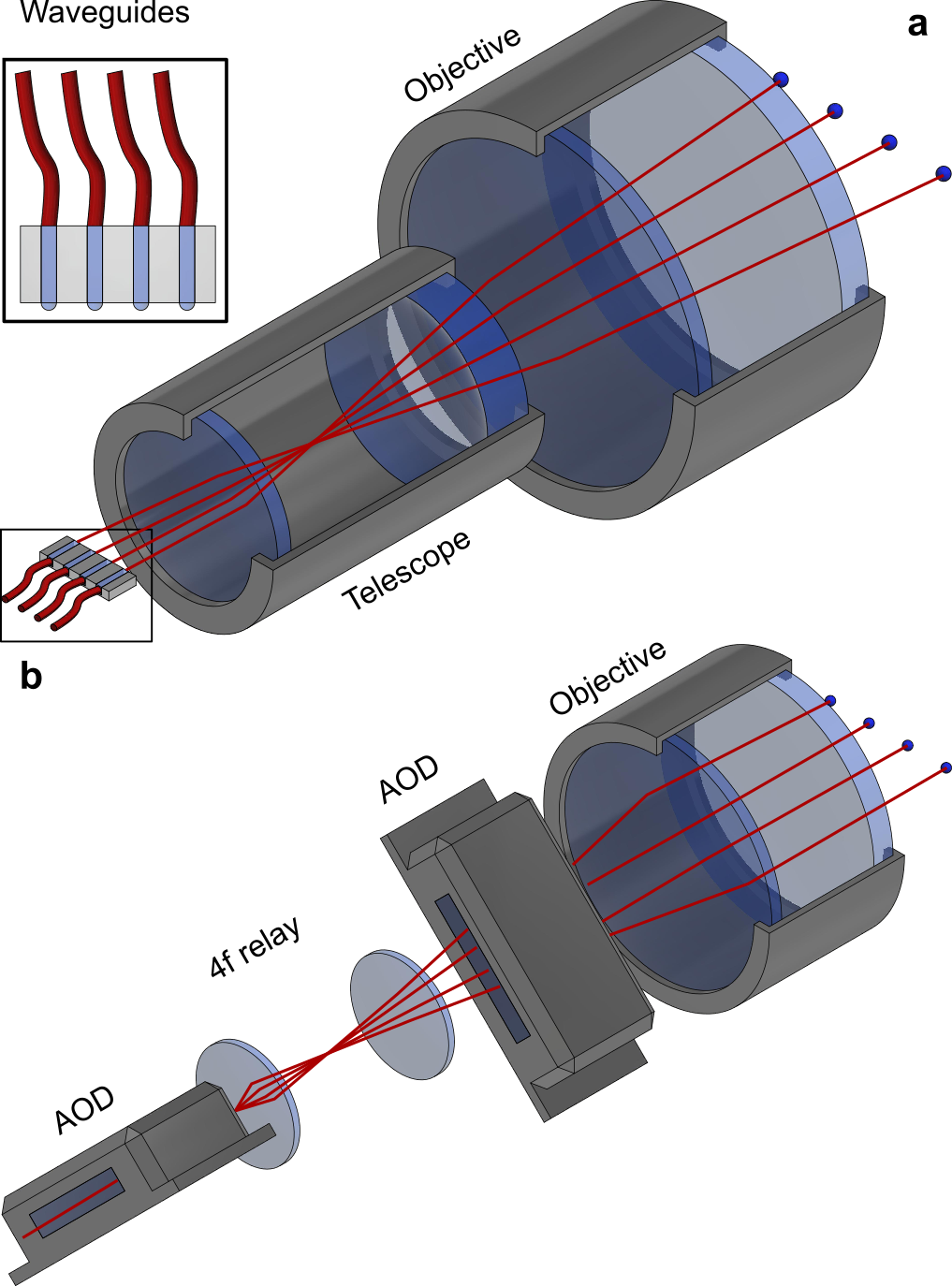}
    \caption{Simplified illustration of addressing approaches trialed, dimensions not to scale. \textbf{a} A fiber-coupled rigid waveguide with microoptics delivers light via a shrinking telescope to the main objective and onto the ions. Beam separations are fixed. \textbf{b} A pair of crossed acousto-optic deflectors (AODs) with a relay telescope delivers light to the main objective and onto the ions. Beam separations are variable.}
    \label{fig:Addressing_Overview}.
\end{figure}

\begin{figure*}
    \centering
    \includegraphics[width=13cm]{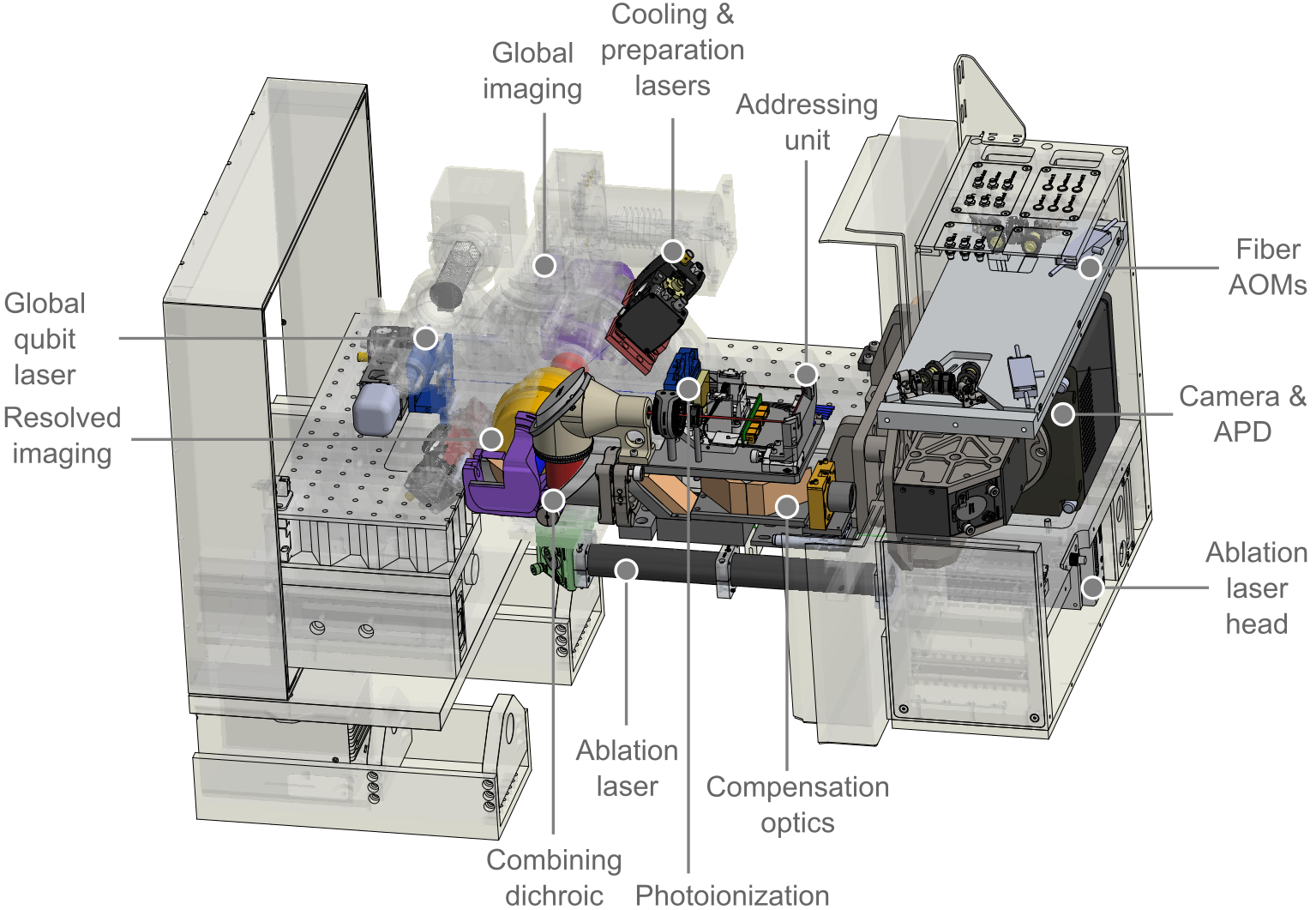}
    \caption{Schematic of trap drawer detail, focusing on optical and detection components, displayed with extended (opened) drawer. Optical interfacing with the ions is achieved via fiber delivery for Doppler cooling, polarization gradient cooling, quenching and repumping (red pair), as well as collective qubit laser (blue, left) and photoionization (blue, right). Free-space delivery of the pulsed ablation laser is achieved via access from below (green). Single-site addressing lasers are steered by an addressing unit, and overlapped with the resolved detection path (high-NA objective, yellow) on a dichroic beam combiner (silver). Compensation optics for resolved light counter abberations introduced via the combiner. Collective imaging to an APD leaves at the back of the chamber (purple). Detection electronics and light sources are situated outside the $\upmu$-metal shielding. Fiber AOMs allow pulsing and imprinting of bichromatic sidebands onto the qubit lasers.}
    \label{fig:TrapDrawer_Optical}
\end{figure*}

\begin{figure}[h!]
		\centering
		\includegraphics[width=8.6cm]{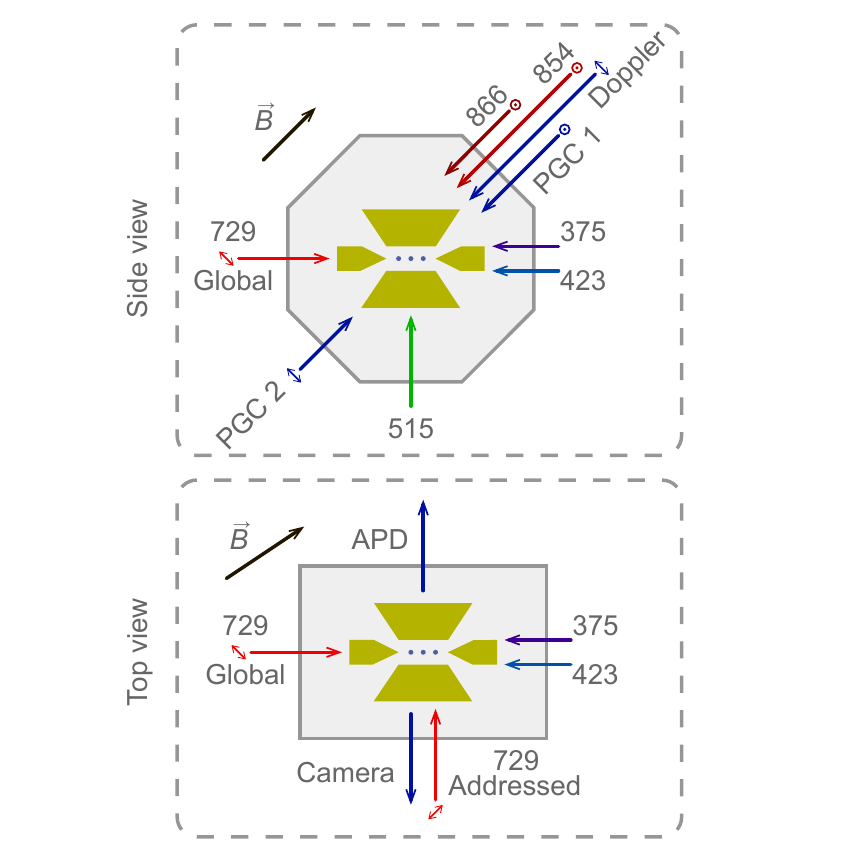}
		\caption[Beam Layout]{Geometric arrangement of beam delivery, beam polarization indicated where required. The Doppler cooling beam, the repumping and quenching beams, as well as PGC beams enter the chamber at 45\degree\ to the trap axis. The collective \SI{729}{\nano\meter} beam is delivered through a hole in the trap endcap. Photoionization beams enter through a hole in the opposing endcap. The medium-NA port is used for collective readout of the whole ion chain with an APD. The high-NA access port is used for single-ion addressing and readout with a camera.}
		\label{fig:BeamLayout}
	\end{figure}

The beamline for the qubit manipulation laser is built analogously, with the addition of active fiber noise cancellation, preceding the switching module. The qubit laser is split on the switching board into two beams before passing a respective free-space double-pass AOM: The first goes directly into a fiber-coupled AOM (FAOM) and is sent to the trap along its symmetry axis for collective operations on all qubits simultaneously. The second is used for single-ion addressing. Light from an addressing unit is delivered to the ions via free space. For the addressing units we trial two different approaches: The first uses a fixed number of fiber-coupled rigid waveguide channels with microoptics which are imaged onto the ion. The second is based on two crossed acousto-optic deflectors (AOD) for positioning of the addressed beam~\cite{Kim:2008}. An overview of these approaches is shown in Fig.~\ref{fig:Addressing_Overview}. The arrangement of light fields arriving at the ion location is shown in Fig.~\ref{fig:TrapDrawer_Optical} and Fig.~\ref{fig:BeamLayout}.

For the microoptics approach, light is sent to an additional splitting module with a fixed number of output channels, followed by individual FAOMs for light switching before being fed to the unit. The FAOMs have a center frequency of \SI{150}{\mega \hertz}, an on/off extinction ratio of $\approx\SI{40}{\dB}$, and imprint the bichromatic sidebands onto the laser, as required for \MS{} interactions. The presence of an individual FAOM per ion has the further benefit of allowing individual control of the light field's parameters for each channel, to e.g. interact with different motional modes for efficient coupling, to adjust each ion's Rabi rate, or to compensate for potential individual qubit frequency shifts. At the stage presented here, the splitting board is composed of a diffractive optic splitting the input light eightfold. The output fibers of the FAOMs are coupled into a waveguide and imaged onto the ion crystal via the main objective lens described in section~\ref{subsec:Imaging}. This addressing unit is provided by Fraunhofer IOF. The details of this unit are beyond the scope of this paper and will be summarized in a forthcoming publication.

In the AOD approach on the other hand light is delivered via a single fiber. The two AODs are oriented at an angle of $\pm45\degree$ with respect to the ion string symmetry axis. Beam steering without incurring position-dependent frequency shifts is achieved by utilizing crossed pairs of AODs where the frequency up-shift from one unit (using +1st diffraction order) is exactly cancelled by the second (using -1st diffraction order). Angular deflection through the AODs is converted to translation in the ion plane by action of the main imaging objective. Driving the AODs with multiple RF tones allows for simultaneous addressing of multiple ions at the cost of additional beams situated above and below the ion string. These beams have a nonzero frequency shift, and their number grows quadratically with the number of RF tones applied. Consequently, the power available per beam decreases quadratically with the number of tones as well. 

\subsubsection{Imaging optics}\label{subsec:Imaging}

Optical readout can be performed in two ways: First, spatially-resolved imaging of the ion string using an EMCCD camera through use of near-diffraction-limited, high-NA objective. Second, not spatially discriminating (collective) light detection through a single-photon-counting avalanche photo diode and medium NA optics. Both detection paths are situated orthogonally to the trap axis and the plane which contains all laser beams to minimize stray light onto the detectors. Additionally, this arrangement gives the largest possible numerical apertures through the vacuum chamber view ports for large collection efficiencies on the imaging systems. Both APD and camera are placed outside of the magnetic shielding, where light exits through penetrations in the shield. Such an arrangement helps to minimize the amount of stray magnetic fields close to the trap chamber which may be caused by the readout electronics in the detectors.

Readout through the APD serves as a means of detection in trap characterization and system diagnostics, but is only suitable for small system sizes, and does not offer site-selectivity. The imaging system employed for APD readout has a medium numerical aperture NA\,$\approx$\,0.29, with a magnification of about $\times1$, and a field of view of about \SI{60}{\micro\meter}. A single ion driven far into saturation on the cycling transition with these parameters yields a photon flux of order 500~kcts/s on the APD given the NA, optical losses and device quantum efficiency. 

The primary imaging system features a high numerical aperture of NA\,$\approx$\,0.5 limited by the trap apertures. This system is used both for site-selective readout, and manipulation of individual ions (single-site addressing). The system's main objective is a compound lens designed to operate near the diffraction limit for the detection light at \SI{397}{\nano\meter} as well as at the addressing light at \SI{729}{\nano\meter}. A dichroic mirror is used to overlap the two colors through this main objective, while compensation optics in the imaging path correct for abberations introduced by the dichroic mirror. Imaging design targets are a magnification of $\approx\times29$ at a field of view of \SI{150}{\micro\meter} to match the detector size. The design modulation transfer function exceeds 0.5 at a resolution of \SI{0.5}{\micro\meter} over the full field of view, well in excess of what is required to resolve single ions with a typical spacing of more $d_0$\,\SI{\geq3}{\micro\meter}.

The main objective's mounting provides five degrees of freedom for precisely aligning the optical axis with the ions necessary to achieving these tight specifications. First, translation perpendicular to the optical axis (X-Y translation) is achieved via flexure joints in the objective's mounting plate, which are actuated via fine-pitch micrometers. Second, pitch and yaw can be adjusted using fine-pitch screws with a spring-loaded kinematic three-point mounting. Piezo-electric actuators allow for fine tuning the position on all three adjusters in addition to coarse alignment via thumbscrews. The final degree of freedom is translation along the optical axis (Z translation or focusing). Coarse alignment is achieved via the fine-pitch threading used to secure the objective lens. This thread is guided through a tight-fitting tube which affords high repeatability. Fine-adjustment is achieved by using the kinematic mount's three piezo-electric transducers in tandem for pure translation. Realignment and refocussing using manual thumb screws is typically only necessary after making major changes to the system. Fine adjustment of the addressing using the piezo actuators is typically done once a day or after opening of the trap drawer.

\subsection{Electronics, control and automation}
Access to and control of the experimental platform is managed by a rack-mounted desktop computer which is accessed via Ethernet. Both racks feature Ethernet switches that connect individual devices to the control computer.

The electronic outfitting of the demonstrator setup is based largely upon modular components. The demonstrator is controlled and driven by both analog and digital electronics. The trap blades providing radial confinement are driven by a dedicated RF signal generator (Rhode \& Schwarz SMB100B) which is amplified via a helical resonator~\cite{siverns2012application}. The ion's secular frequency is actively stabilized by a feedback circuit actuating on the supplied trap RF~\cite{johnson2016active}. In addition, the demonstrator makes use of state-of-the-art phase-coherent digital signal generation, real-time control, and in-sequence decision making based on field programmable gate arrays (FPGAs) to perform digital (pulsed) operations.

\subsubsection{Experimental control electronics}
\begin{figure*}
    \centering
    \includegraphics[width = 13.07cm]{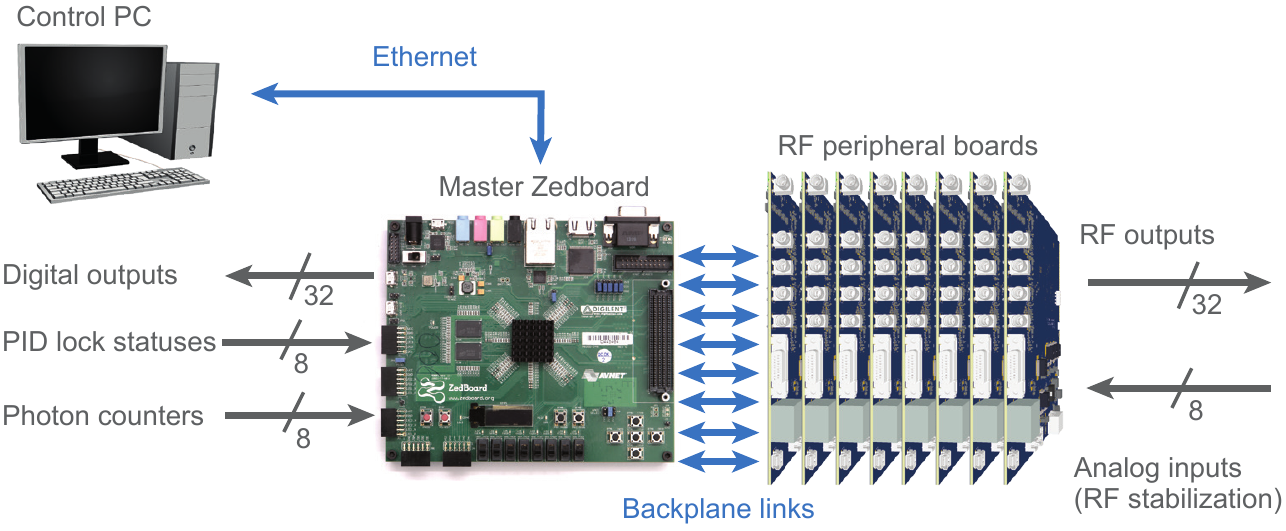}
    \caption{M-ACTION experimental control system. The master board acts as the hub of the system, synchronously running experimental sequences involving the digital I/O and the RF peripheral boards, and managing communications with the control PC.}
    \label{fig:MACTION}
\end{figure*}

The laser pulses used for manipulating the states of the ions are controlled using acousto-optic modulators.
These require radio frequency signals which are precisely timed and phase-coherent (inter-pulse, inter-channel, and inter-module). For typical RF pulses several microseconds in length, the timing resolution needs to be \SI{<10}{\nano\second} to control pulse lengths to better than \SI{1}{\percent}, and the timing jitter must be below a few hundred picoseconds to ensure repeatability.
Digital input/outputs (I/Os) with similar timing resolution and jitter are also required, both for triggering external devices such as RF switches, shutters or arbitrary waveform generators, and for counting photon-arrival pulses from photomultipliers or APDs.
These RF and digital signals are manipulated using an FPGA-based modular experimental control system developed at ETH Zurich, known as `Modular Advanced Control of Trapped IONs' (M-ACTION)~\cite{negnevitsky2018feedback}.
It uses a 'star' topology as shown in Fig.~\ref{fig:MACTION}, with a central master control board that communicates with the control PC via a custom Ethernet protocol, and multiple peripheral boards providing RF output~\cite{keitch2017programmable}. The peripheral boards communicate with the master via a low-level low-voltage differential signaling (LVDS)-based protocol through a backplane into which all the boards are inserted.

The master board is a commercially-available Avnet Zedboard extended with custom boards for digital I/O and board-to-board communication.
It is centered around a Xilinx Zynq system-on-chip, which holds both an FPGA core used for low-level deterministic signal handling, and a dual-core \SI{667}{\mega\hertz} ARM A9 CPU suitable for higher-level control and real-time computations. These include Boolean decisions such as those in quantum error correction, as well as more sophisticated Bayesian decisions featuring feedback of a continuous parameter to the experimental hardware~\cite{negnevitsky2018feedback,DeNeeve:2017}.
The FPGA core, on the other hand, monitors the peripheral RF boards, and controls the digital I/O that requires precise timing.

The RF peripheral boards each consist of four direct-digital synthesis (DDS) chips (Analog Devices AD9910), controlled by a standalone FPGA (Xilinx Spartan-6 XC6S150T) with sufficient memory and a pulse sequence processor to independently execute most typical experiments without intervention from the master~\cite{keitch2017programmable}. 
In situations where intervention from the master board is required, such as when real-time feedback is being carried out, the latency between an input to the master and a change in pulse sequence on the peripheral cards is around \SI{5}{\micro\second}.
Recent revisions of the RF board also feature two analog-to-digital converters, and support on-board servo regulators for intensity stabilization with sample and hold synchronous to the local pulse sequences. Regulator gains and setpoints may be altered in-sequence to help cope with external nonlinearities.

The aforementioned RF boards support a single-tone output per channel, requiring multiple channels to be added together for multi-tone quantum gates. Additionally, the complexity of a pulse is limited by the bandwidth of the interface between the FPGA and the DDS chips. A new RF generation board using 1~GSa/s DACs controlled directly by a more powerful FPGA, as well as 32~GB of local memory, is being currently developed to overcome these bottlenecks.

In-sequence decision making is performed based on pre-determined conditional decisions or 'branches'. This pre-determination allows FPGA memory management. Branches are sub-sequences that are executed if in-sequence detection fulfills the branch conditions. Photon count data from either a camera or APD is used directly on the FPGA to determine whether a branch condition is met.

A simple example would be of a state transfer $\ket{S}\to\ket{D}$ by means of a $\frac{\pi}{2}$ pulse. If at the decision making point $\ket{S}$ is measured, that is many photons are scattered in the detection window, then the FPGA determines that the condition for state transfer to $\ket{D}$ is met, and executes a $\pi$-pulse to transfer the population into $\ket{D}$. If on the other hand $\ket{D}$ is detected, that is few photons are scattered, this additional pulse is not executed. A more complicated example is given in the literature~\cite{stricker2020experimental}, which uses the same hardware in a separate experiment.

\subsubsection{Experimental control software}
The experiment hardware is operated using two layers of control software.
The lower layer, running on the master board, is written in C$++$ and provides access to the functionality of M-ACTION to a Python-based higher layer running on the control PC.
This higher layer is used for day-to-day experimental operation and programming, while the lower layer can be used to implement high-performance real-time computations, or extend the capabilities of M-ACTION.

A graphical user interface (UI) gives direct feedback to the operator, and is used for the majority of day-to-day experimental control tasks. Script-based implementations of pulse sequences allow for versatile extensions to low-level pulse patterns available from the UI.
Automated routines for calibration, maintenance and error handling, such as Rabi frequency tracking, frequency drift compensation, and string recrystallization, are supported to maintain the system at high fidelity without constant supervision.
Quantum circuits are executed via a high-level language made in house, called PySeq, which acts as an intermediary between state-of-the-art frameworks like Cirq~\cite{Cirq} and Qiskit~\cite{Qiskit} and close-to-hardware descriptions on the laser pulse level. PySeq is, similar to Qiskit, a Python package and provides a convenient abstraction layer translating quantum circuit objects like rotations around an axis for a given angle to hardware instructions such as laser pulses with a certain power and duration.
It is expected that this feature, in combination with remote access to the setup, will greatly improve the ease of use for collaborators.

\section{System characterization}
\label{sec:Characterization}
We now turn to characterizing the demonstrator setup. A suite of common measures and experiments is used to evaluate the engineering approaches for the rack-based architecture. The section begins with outlining standard operation parameters for the ion trap. Measurements on the mechanical stability of the rack and the performance of the active vibration isolation system are presented. We show measurements on imaging system performance, as well as addressing capabilities using two different addressing unit devices. We finally turn to measurements pertaining to the operation of the demonstrator itself, such as characterization and compensation of the magnetic field gradients, coherence times, ion temperature and heating rates, and gate performance for single and pairwise qubit interaction.

\subsection{Trapping parameters}

The Paul trap's radial confinement voltage is supplied to a helical resonator, in order to filter and impedance match the RF drive to the trap~\cite{siverns2012application} which results in a voltage step-up at the trap side. The loaded resonator oscillates at a drive frequency of $\Omega_\textrm{RF} \approx 2 \pi \times$ \SI{27.4}{\mega\hertz}. It is driven with an input power of up to $P_\textrm{in,max} =$ \SI{10}{\watt}. The trap endcaps are supplied with a voltage of up to $V_\textrm{ec} \approx$ \SI{1000}{\volt}. An ion trapped in the trap will oscillate at the resulting pseudopotential's three fundamental trap frequencies $\omega_\text{ax}$, and $\omega_\text{rad}$ which can be measured by external excitation with oscillating fields ('tickling')~\cite{dholakia1992photon} or by sideband spectroscopy~\cite{mavadia2014optical}. These trap frequencies are (nearly) identical to the center-of-mass motional excitations of ion clouds or crystals in the three trap directions. We determine the secular frequencies for a single \cai of $\omega_\textrm{rad} \approx 2 \pi \times$ \SI{3}{\mega\hertz} in the radial direction with \SI{10}{\watt} input power, and  $\omega_\textrm{ax} \approx 2 \pi \times$ \SI{1}{\mega\hertz} along the trap axis at about \SI{1000}{\volt} endcap voltage. In situ temperature measurements using a PT100 thermistor give a temperature of the trap that depends on RF drive power. The temperature reaches about $T_\textrm{max} \approx $ \SI{100}{\celsius} after sustained operation at $P_\textrm{in,max} =$ \SI{10}{\watt}. The pressure in the vacuum chamber as measured by the pump's gauge is \SI{1.5e-11}{\milli\bar}. The pressure reading does not depend on the drive power, indicating little outgassing from adsorbed contaminants on the trap surfaces.

The ion gauge's pressure measurement is performed at the pump location instead of the ion location. Differential pressures in ultra-low vacuum conditions can thus lead to an underestimation of the pressure at the ion location. Consequently, we independently determine the pressure at the ion location via collision rate measurements~\cite{hempel2014digital,furst2019trapped}. Collisions with residual thermal atoms melt the ion crystal and thus limit the stability of ion crystals in the trap. We distinguish between two different types of collision events: Those that melt the crystal and cause ions to change sites in the refrozen crystal, and those that lead to molecule formation via chemical reactions which changes the number of (bright) ions. Detection of reconfiguration events is done by observing the position of non-fluorescing, co-trapped ions (defect hopping).  We measure a collision rate of $\Gamma_\textrm{col} =$ \SI{0.0025(11)}{\per\second} per ion by observing reconfiguration events in a three ion crystal consisting of two bright \cai and one dark, co-trapped ion, which corresponds to a pressure of $p_\textrm{ion} = \SI{9.7(42)e-11}{\milli\bar}$ assuming collisions with H$_2$~\cite{furst2019trapped}. This means that even a 50-ion chain is unlikely to experience a collision during the qubit coherence time of \SI{100}{\milli\second}. Ion loss may occur when collisions are sufficiently strong to lead to unstable ion trajectories, or when chemical reactions cause formation of dark ions. By observing an ion crystal of 32 bright \cai ions over the course of \SI{12}{\hour} we observe the loss of only one ion from the trap.

\subsection{Mechanical stability and active damping}

\begin{figure}[h!]
		\centering
		\includegraphics[width=8.6cm]{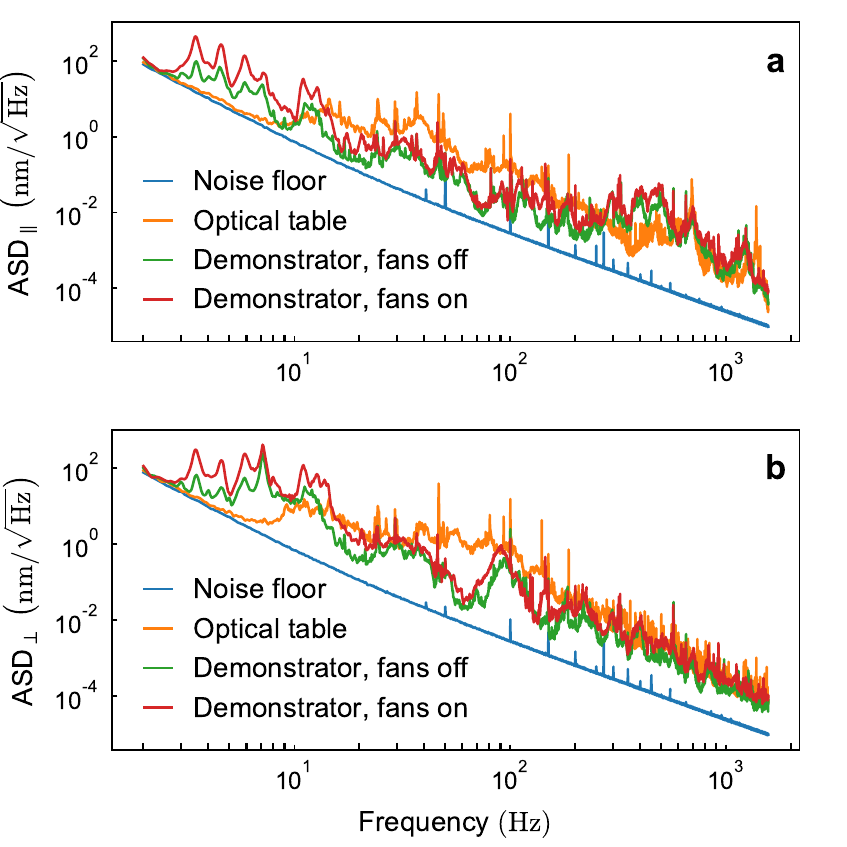}
		\caption{Displacement (amplitude) spectral densities (ASD) \textbf{a} parallel to the floor (horizontal) and \textbf{b} perpendicular to the floor (vertical) calculated from time series data obtained with piezo-electric shear accelerometers. Traces show equivalent displacement spectral density of the electronic noise floor (blue), and spectral densities of typical optical table setup in close proximity (orange), demonstrator setup with cooling fans turned off (green) and on (red).}
		\label{fig:Vibrations}
	\end{figure}

Mechanical instability affects quantum computing fidelity via a multitude of avenues from light field amplitude fluctuation, lowering of signal-to-noise through blurring or defocusing in the detection system, to light field dephasing noise~\cite{micke2019closed, kessler2012sub, hartnett2012ultra, somiya2012detector, aguirre2015dark}. We measure the vibrational stability of the setup using piezo-electric shear accelerometers (Endevco 7703A-1000) at various locations in the setup. Time series data from the detectors is recorded and Fourier transformed to voltage power spectral densities in an audio spectrum analyzer (Stanford Research Systems SR1). Voltage spectra are converted to displacement spectral densities via the accelerometer sensitivity calibration. The measurement is limited by the electronic noise floor of the signal conditioner (Endevco 133) below about \SI{2}{\hertz}. Displacement spectral densities recorded simultaneously in the horizontal and vertical directions are shown in Fig.~\ref{fig:Vibrations}, along with a reference spectrum obtained on an empty, broadband-damped, honey comb-lattice optical table in close proximity. Root-mean-square (RMS) displacements over the full measurement bandwidth are shown in Tab.~\ref{Tab:RMS} with electronic background subtracted. 

\begin{table}[h!]
    \centering
    \begin{tabular}{c|c|c|c|c}
          & Noise & Opt. table & Full w/o fans & Full w/ fans  \\\hline
         RMS$_\text{hor.}$ (nm) & 54 & 21 & 61 & 275\\
         RMS$_\text{vert.}$ (nm)  & 50 & 33 & 139 & 335\\ 
    \end{tabular}
    \caption{Root-mean-square (RMS) displacements over the full measurement bandwidth for vibration measurements shown in Fig.~\ref{fig:Vibrations}. Electronic background subtracted point-wise from spectra.}
    \label{Tab:RMS}
\end{table}
	
The overall vibrational stability of the system is similar to traditional optical setups situated on dedicated optical tables. Spectra do show a clear influence of the operation of cooling fans used in the rack construction. The majority of root-mean-square (RMS) displacement is contributed at low frequencies below \SI{10}{\hertz}. Notably these frequencies are also well below the fan rotation frequencies, indicating that they do not simply originate from vibrations of the fans or modulations at the \SI{50}{\hertz} motor drive. Determining the origin of excess vibrations caused by the fans and the structured noise at low frequencies is part of future upgrade efforts, including DC-drive fans and improved air flow direction. The performance is comparable to the empty reference table in spite of the compact, tower-like construction in active operation, and validates engineering approaches used in the demonstrator.

\subsection{Imaging and detection performance}

\begin{figure*}
    \centering
    \includegraphics[width=15cm]{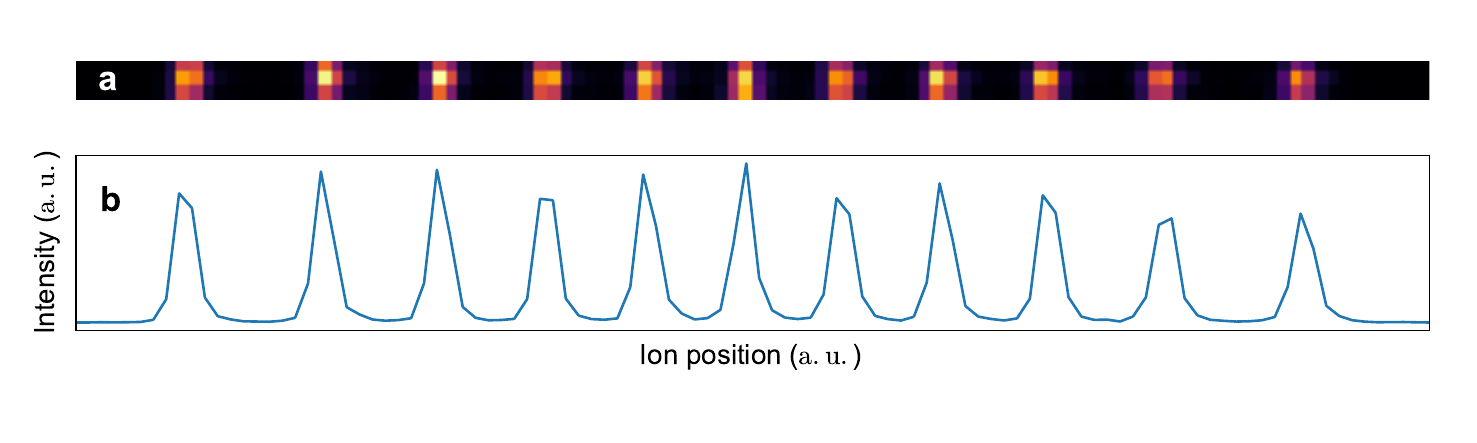}
    \caption{Detection performance. \textbf{a} Fluorescence image of a linear string containing 11 ions. The axial trap frequency is $\omega_\text{ax} = 2\pi\times \SI{450}{\kilo\hertz}$, corresponding to a distance between the two center ions of \SI{4}{\micro\meter}. \textbf{b} Integration over pixel columns shows well-separated ions.}
    \label{fig:Imaging}
\end{figure*}

Characterizing the performance of the primary imaging system provides information about the achievable signal-to-noise ratio in readout, and the minimal achievable spot size and crosstalk for addressing light. After optimization of the objective alignment we resolve well-separated ions in a 50-ion linear crystal. A typical image of an 11-ion crystal with minimal ion-ion separation of $\approx \SI{4}{\micro\meter}$ is shown in Fig.~\ref{fig:Imaging} \textbf{a} recorded at the detection wavelength of \SI{397}{\nano\meter}. We determine the achieved imaging system magnification to be $\times 29.9$ from the known pixel dimensions and calculated ion spacings for the used trap parameters. This is slightly larger than the design magnification of $\times29$. The high-NA readout optics enable camera detection times down to \SI{300}{\micro\second} at a fidelity of about 99.9\,\%.

We determine the detection fidelity by preparing an ion string either in the bright 4\,S$_{1/2}$, or in the dark 3\,D$_{3/2}$ manifold by turning off the repumping laser at \SI{866}{\nano\meter}.  We elect to pursue this method over creation of the dark state via excitation using the \SI{729}{\nano\meter} laser to avoid conflating operational infidelities in qubit manipulation with detection infidelities. After a cooling period we perform a projective measurement and detect each ion individually on a camera. From the counting statistics in each ion detection region we determine the probability of measuring the prepared state for a given measurement time, that is, measure many (no) photons for the ion prepared in the bright (dark) state. We estimate the detection crosstalk by determining the probability of measuring a bright state in a detection region in between two bright ions spaced at \SI{5.9}{\micro\meter}. Detecting light above the background in this empty region of space comes solely from detection crosstalk. At a measurement time of \SI{300}{\micro\second} we detect no events above the background threshold out of 10000. Consequently, we conclude that detection crosstalk during our measurement windows is negligible.


\subsection{Single ion addressing performance}
\label{subsec:Addressing}
Addressing capabilities are crucial for the quantum computing scheme we pursue. This capability is routinely quantified in terms of the ability to resolve individual sites along with the crosstalk between individual channels. In the following, characterization measurements of the microoptics addressing unit and AOD addressing units are presented.

\begin{figure*}
    \centering
    \includegraphics[width = 15cm]{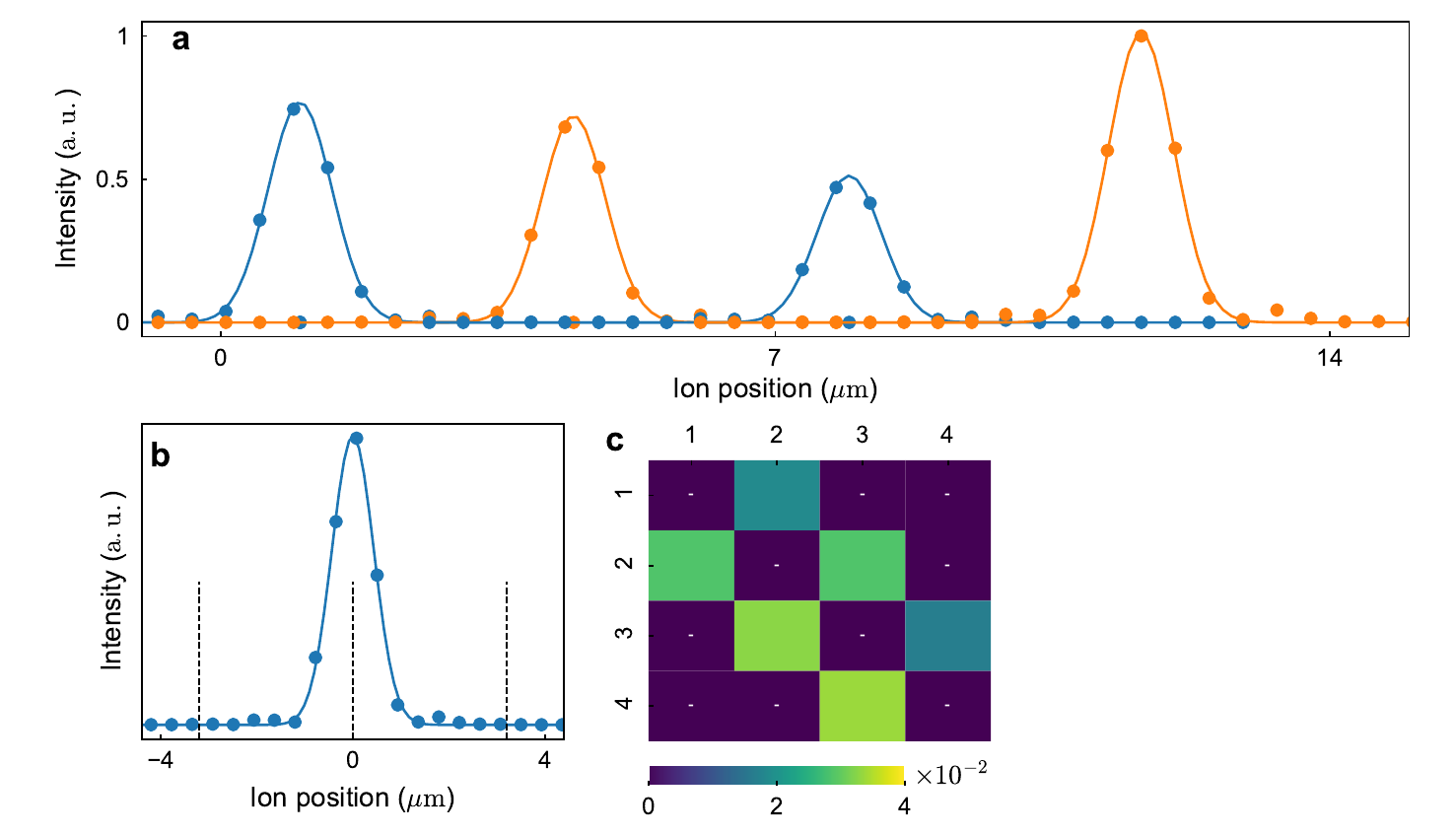}
    \caption{Microoptics addressing unit performance. \textbf{a} A single ion's position is scanned electrostatically along the trap axis by varying the endcap voltages. The square of the Rabi frequency is plotted as a function of this position for four active channels at a distance of \SI{3.4}{\micro \meter}. \textbf{b} Same as \textbf{a} with higher resolution for spot size measurement of the focused addressing beam at \SI{729}{\nano\meter}. By fitting a Gaussian distribution to the data we obtain a beam width of about $w_0 \approx \SI{0.8}{\micro\meter}$. Dashed black lines indicate the nearest neighbors for a 50-ion crystal at the minimal distance of $\approx\SI{3.2}{\micro\meter}$. \textbf{c} Estimates of resonant crosstalk from intensity profiles in \textbf{a}. Entries with '-' were not assessed.}
    \label{fig:Addressing_IOF}
\end{figure*}

At this stage we utilize only four of the channels on the microoptics addressing unit. The beams from these channels are fixed in position, and measuring the beam profile proceeds by moving a single ion electrostatically along the trap axis through the beams by scanning the endcap voltages. At each position the light intensity is obtained by measuring the oscillation frequency of Rabi oscillations. The Rabi frequency is proportional to the electric field at a given position. Figure~\ref{fig:Addressing_IOF} \textbf{a} shows the estimated light intensity along the trap axis obtained in this fashion. From a Gaussian fit to a higher-resolution scan shown in Fig.~\ref{fig:Addressing_IOF} \textbf{b} we calculate a beam waist of $w_0 = \SI{0.81(1)}{\micro\meter}$ for all spots. Different coupling efficiencies through FAOMs and waveguides lead to peak heights varying across channels.

We further measure the resonant crosstalk between channels as defined by the ratio of the Rabi frequencies, $\Omega_\textrm{adjacent}/\Omega_\textrm{addressed}$ when exciting a single ion. For the four active channels we measure resonant crosstalk between \SI{1.9}{\percent} and \SI{3.4}{\percent} as shown in Fig.~\ref{fig:Addressing_IOF} \textbf{c}. This is much larger compared to what we would expect from a Gaussian beam with $w_0 = \SI{0.81}{\micro\meter}$ or Airy fringes arising from the finite aperture, and implies that the crosstalk is limited by optical aberrations. 

For the AOD addressing unit we scan the addressing beam over a string of ions by changing the AOD drive frequency. In this measurement we use laser pulses of a fixed power and duration such as to produce a pulse area of $<\pi$ at the center of the beam. Scanning the beam across an ion will thus produce Rabi oscillations with maximal population in the excited state coinciding with maximal intensity. We calibrate the deflection per drive frequency increment obtained from the AODs by comparing the population profile to the known ion distances at a given axial center-of-mass mode frequency. The calibration yields a deflection slope of \SI{4.9(1)}{\micro \meter \per \mega \hertz}.

\begin{figure*}
    \centering
    \includegraphics[width = 15cm]{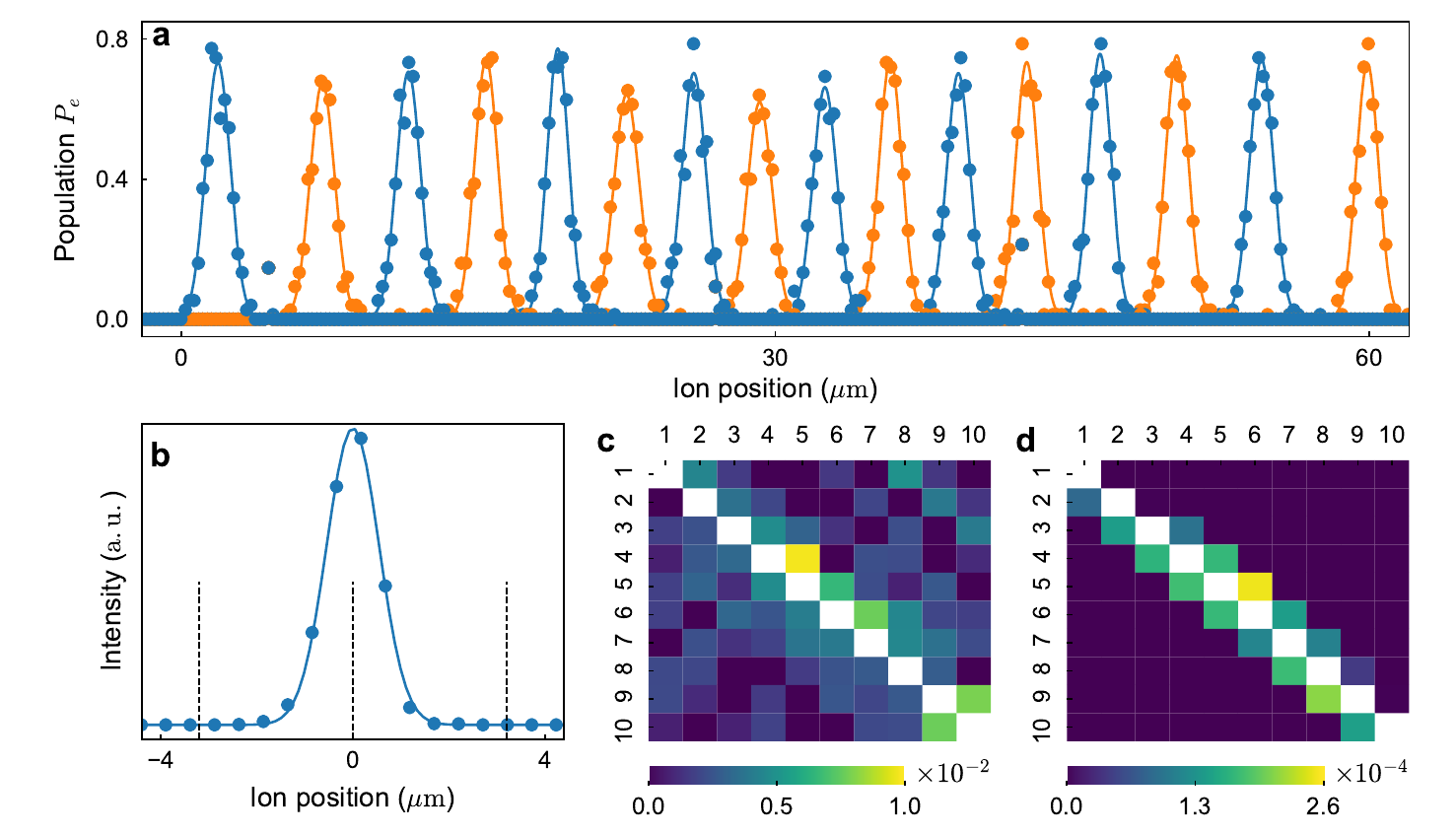}
    \caption{AOD addressing unit performance. \textbf{a} Scan of addressing beam over a 16-ion crystal. The population $P_g$ is used as a proxy for signal intensity, see main text. \textbf{b} Spot size measurement of the focused addressing beam at \SI{729}{\nano\meter}. A single ion's position is scanned electrostatically along the trap axis by varying the endcap voltages. The square of the Rabi frequency is plotted as a function of this position. By fitting a Gaussian distribution to the data we obtain a beam width of about $w_0 \approx \SI{1.1}{\micro\meter}$. Dashed black lines indicate the nearest neighbors for a 50-ion crystal at the minimal distance of $\approx\SI{3.2}{\micro\meter}$. \textbf{c} Resonant crosstalk measurements in 10-ion crystal with minimal inter ion distance of $\SI{3.5}{\micro \meter}$. Plotted are the ratios of Rabi frequencies of ions relative to the addressed ion (white on diagonal). \textbf{d} Same as \textbf{c} but using off-resonant, compensating $U(3)$ pulses, see main text. Ions where Rabi oscillations are too slow to fit reliably are set to 0 in the plot to avoid spurious structure.}
    \label{fig:Addressing_AOD}
\end{figure*}

A measurement for a 16-ion string with a minimal ion distance of $\approx\SI{3.4}{\micro\meter}$ and a total length of $\approx\SI{60}{\micro\meter}$ is shown in Fig.~\ref{fig:Addressing_AOD} \textbf{a}. The intensity profile shown in Fig.~\ref{fig:Addressing_AOD} \textbf{b} is measured analogously to the microoptics addressing unit's, yielding a beam waist of $w_0 \approx \SI{1.09(2)}{\micro\meter}$. Next, we characterize the resonant and off-resonant crosstalk in a 10-ion string with minimal inter-ion distance of $\SI{3.5}{\micro \meter}$. The ratio of the Rabi frequencies of the non-addressed ions to the addressed ion is plotted in Fig.~\ref{fig:Addressing_AOD} \textbf{c}. We obtain an average resonant crosstalk of \SI{0.2}{\percent} with a maximum of \SI{1}{\percent} on ion 5 when addressing ion 4. The average nearest-neighbor crosstalk is \SI{0.5}{\percent}. This is significantly lower than in the microoptics addressing unit. The difference may at least partly be attributed to the better overall optical quality of macroscopic precision optics used in the AOD approach, rather than microoptical prototype components.

Figure~\ref{fig:Addressing_AOD} \textbf{d} shows the results of off-resonant crosstalk measurements. For this measurement we perform a resonant collective $\pi/2$-pulse, followed by an addressed AC-Stark pulse of variable length, followed by a collective $-\pi/2$-pulse. These composite pulses, sometimes referred to as $U(3)$ pulses, are used to reduce the effect of crosstalk. The AC-Stark shift is proportional to the light field intensity $I\propto E^2$ while the Rabi frequency is proportional to the electric field $\Omega \propto \sqrt{I} \propto E$. Consequently, we expect the $U(3)$ pulses to produce significantly smaller crosstalk. The maximum crosstalk measured is $2.6\times 10^{-4}$, thus 20 times lower compared to the resonant crosstalk. The average crosstalk on nearest-neighbors is $1.3\times10^{-4}$.

\subsection{Coherence properties and magnetic fields}	

Qubit coherence times are affected by a multitude of sources. Two common technical sources of decoherence in state-of-the-art trapped-ion setups are phase noise from the driving fields, and magnetic field noise. Phase noise may originate either directly from laser frequency instability, that is finite linewidth, or can be imparted through e.g. fiber noise or optical path length noise. Magnetic field noise modulating the energy splitting in the qubit manifold often originates from mains or its harmonics, switching power supplies, or ground loops.

Phase noise in the driving field is ultimately limited by the linewidth of the stabilized laser. Vibrational isolation, rigid construction principles, fixing of flexible (fiber optics) paths and air current control in free-space are implemented throughout the setup to prevent adding excess phase noise. Additionally, fiber noise cancellation is used up to the double pass switching AOM, see section~\ref{subsec:LightDelivery}.

The influence of magnetic field noise depends on two factors: The noise present and the sensitivity of the utilized transition to it. The qubit transition $\ket{\textrm{4\,S}_{1/2},m_J = -1/2} \leftrightarrow \ket{\textrm{3\,D}_{5/2},m_J = -1/2}$ is the least sensitive to magnetic field fluctuations out of the $\textrm{S}_{1/2} \leftrightarrow \textrm{D}_{5/2}$ manifold, with a slope of \SI{5.6}{\mega\hertz\per\milli\tesla}. This is orders of magnitude higher than in clock transitions of hyperfine qubits~\cite{brown2018comparing}, and therefore magnetic field stability as well as homogeneity have to be precisely controlled in order to reach sufficient coherence times on the order of \mbox{$T_2 \approx$ \SI{100}{\milli\second}} or better.

\begin{figure*}
    \centering
    \includegraphics[width=15cm]{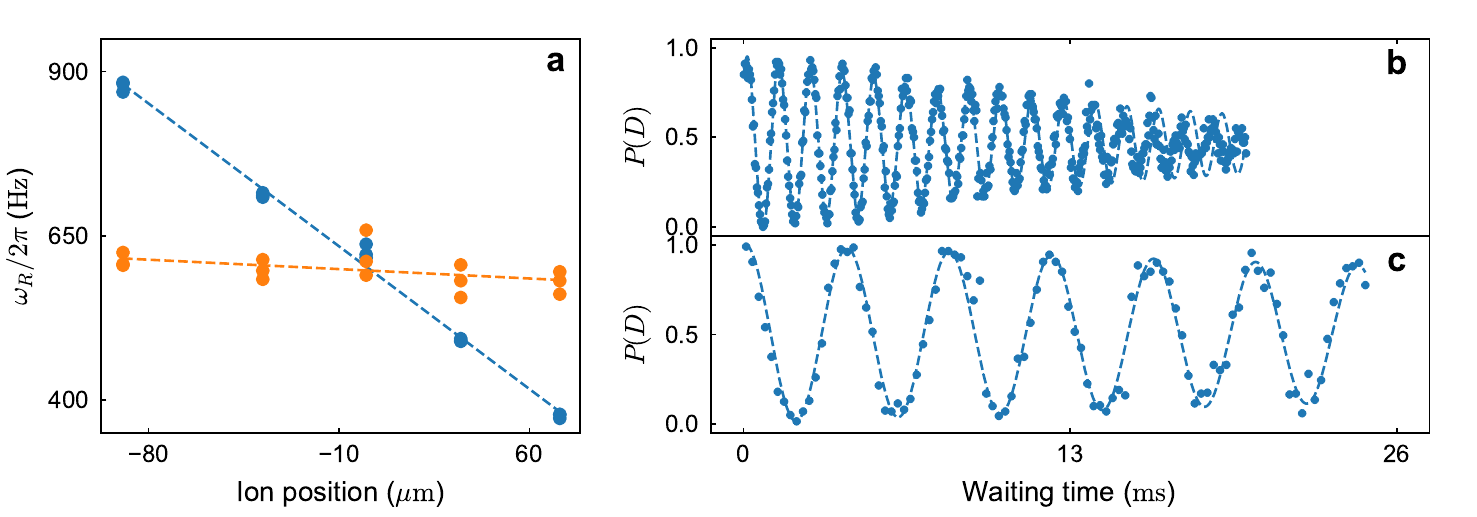}
    \caption{\textbf{a} Magnetic field gradient before compensation (blue, \SI{3.2(1)}{\hertz\per\micro\meter}) and after compensation (red, \SI{0.2(1)}{\hertz\per\micro\meter}) on ground state qubit as seen from Ramsey fringe frequency $\omega_R$ for a fixed detuning as a function of ion position, corrected for linear frequency drift at fixed position.   \textbf{b} Decay of Ramsey fringes of dark state population $P(D)$ for the ground state qubit over time, resulting in a coherence time of \mbox{$T_2^{\text{GS}} =$ \SI{18(1)}{\milli\second}}. \textbf{c} Same for optical qubit transition. The sensitivity to magnetic fields is five times lower, leading to less impact from magnetic field noise. We determine \mbox{$T_2^{\text{Opt}} =$ \SI{90(30)}{\milli\second}}.}
    \label{fig:Coherence}
\end{figure*}

We characterize the temporal magnetic field stability by performing Ramsey spectroscopy on the \textit{ground state qubit} $\ket{\textrm{4\,S}_{1/2},m_J = -1/2} \leftrightarrow \ket{\textrm{4\,S}_{1/2},m_J = +1/2}$ transition in a single ion. This transition has a five times increased sensitivity to the magnetic field with respect to the qubit transition, with a gradient of \SI{28.0}{\mega\hertz\per\milli\tesla}. The optical pulse sequence to access this transition begins with a $\pi/2$-pulse, which creates an equal superposition state between $\ket{0}$ and $\ket{1}$. A $\pi$-pulse then coherently transfers the population from the optical qubit $\ket{0}$ state to the $\ket{\textrm{4\,S}_{1/2},m_J = +1/2}$ state. After the Ramsey waiting time the sequence is reversed, which implements the Ramsey time evolution of an effective magnetic dipole transition. This sequence yields a significant decrease in sensitivity to optical noise sources, thus isolating the influence of the magnetic field. Performing Ramsey spectroscopy with closed $\upmu$-metal shield on the ground state qubit and optical qubit yields respective coherence times of \mbox{$T_2^{\text{GS}} =$ \SI{18(1)}{\milli\second}} and \mbox{$T_2^{\text{Opt}} =$ \SI{90(30)}{\milli\second}}, as shown in Fig.~\ref{fig:Coherence} \textbf{b} and \textbf{c}. This is sufficient to execute about 400 two-qubit gates or 4000 single-qubit gates during the coherence time.

Furthermore, we measure no influence of the vibration isolation elements on the coherence time of our qubit. The initial decay of fringe contrast is consistent with coherence times in excess of the \mbox{$T_2^{\text{Opt}} =$ \SI{90(30)}{\milli\second}}. However, for waiting times longer than \SI{25}{\milli\second} we observe accelerated contrast decay which limits the coherence time to the value above, and causes the substantial uncertainty. The source of this enhanced decay is the subject of ongoing investigation, but may be caused by magnetic field noise induced by switching magnets in neighboring experiments, or mains noise. This is further supported by two observations: First, neighboring ion trap experiments implementing feedforward to cancel mains noise see a substantial improvement in coherence times. Second, a similar experiment being set up in a neighboring building with overall worse environmental control but almost no other occupants causing magnetic field noise also shows increased coherence times.

In addition, the spatial homogeneity of the magnetic field needs to be investigated to maintain coherence across an extended ion register. Small deviations in the positioning of the employed Halbach and Helmholtz configurations may cause a gradient or curvature across the trap axis. Likewise, the permanent magnet of the ion getter pump may cause a magnetic field gradient across the ion string. We measure a gradient $\partial_z B =$ \SI{0.111(2)}{\milli\tesla\per\meter} by performing Ramsey spectroscopy on the ground state qubit and then shifting the equilibrium position of a single ion electrostatically. The gradient leads to a frequency shift on the ground state qubit transition of \SI{3.1(1)}{\hertz\per\micro\meter}. Applying appropriate currents to the compensation coils reduces this axial gradient by an order of magnitude to \SI{0.2(1)}{\hertz\per\micro\meter}. For the five times less sensitive optical qubit transition this means a gradient of \SI{0.041(21)}{\hertz\per\micro\meter}, corresponding to an end-to-end frequency difference of $\approx\SI{8}{\hertz}$ in a \SI{200}{\micro\meter} chain of 50 ions, as shown in Fig.~\ref{fig:Coherence} \textbf{a}.

\subsection{Ion temperature and motional heating rates}
	
\begin{figure}[h]
    \centering
    \includegraphics[width=8.6cm]{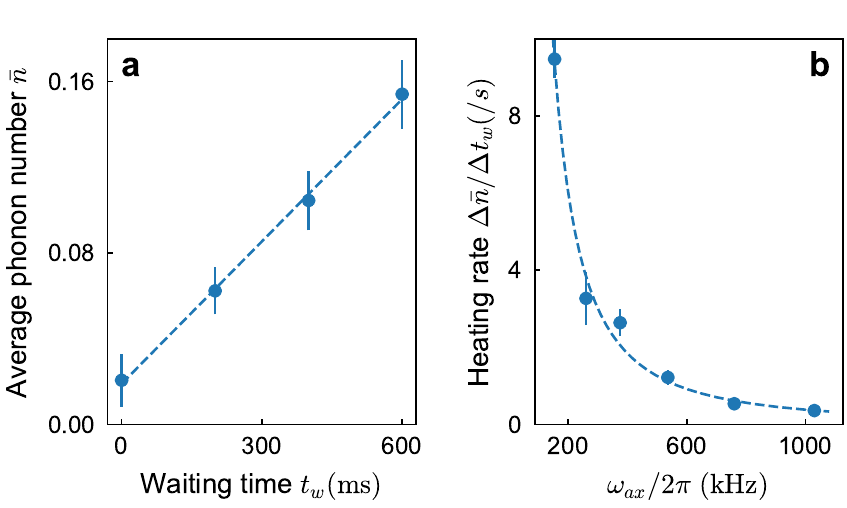}
    \caption{Axial heating rates. \textbf{a} We calculate the heating rate of $\Delta\bar{n}/\Delta t_w =\SI{0.221(7)}{\per\second}$ at $\omega_\textrm{ax} = \SI{1.05}{\mega\hertz}$ by fitting the increase in $\bar{n}$ over the waiting time $t_w$, where $\bar{n}$ is determined from sideband thermometry. \textbf{b} Heating rates are determined as in \textbf{a} as a function of trap frequency. The dashed line is a power-law fit $1/\omega_\text{ax}^\alpha$, where $\alpha\approx1.7$.}
    \label{fig:HeatingRate}
\end{figure}

High fidelity gate operations typically require the ions to retain coherence in their motional degrees of freedom~\cite{turchette2000decoherence,brownnutt2015ion,talukdar2016implications,Ballance:2016}. Motional coherence is ultimately limited by the motional heating rate. We therefore measure both the ion temperature, that is phononic mode occupation, and from that the heating rates with sideband thermometry~\cite{epstein2007simplified}. Sideband thermometry can be applied for an average phonon occupation $\bar{n}\lesssim 2$. Comparison of the Rabi oscillations on the red and blue sideband of the qubit transition yields the average phonon occupation ~\cite{haffner2008quantum}. After sideband cooling a single ion, we obtain a phonon number of $\bar{n}_\textrm{ph, ax} =$ \SI{0.02(1)}{} (at $\omega_\textrm{ax} = 2\pi\times\SI{1.05}{\mega\hertz}$) in the axial direction, and $\bar{n}_\textrm{ph, rad} =$ \SI{0.06(2)}{} (at $\omega_\textrm{rad} = 2\pi\times\SI{2.5}{\mega\hertz}$) in the radial direction. By increasing the time between sideband cooling and temperature measurement we obtain a heating rate of \SI{0.221(7)}{\per\second} in the axial direction (see Fig.~\ref{fig:HeatingRate} \textbf{a}, and \SI{0.3(1)}{\per\second} in the radial direction. These heating rates compare favourably with values obtained in other traps operated at room temperature~\cite{Brownnutt:2015} (see in particular Fig.~8). Measuring heating rates for axial trap frequencies between $\omega_\textrm{ax} = 2\pi\times\SI{0.15}{\mega\hertz}$ and $\omega_\textrm{ax} = 2\pi\times\SI{1.05}{\mega\hertz}$, we obtain a power-law dependency $1/\omega_\text{ax}^\alpha$, with $\alpha\approx1.7$, see Fig.~\ref{fig:HeatingRate} \textbf{b}.
	
\subsection{Single qubit gates}
Arbitrary single-qubit rotations are part of the complete gate set we chose to implement for universal quantum computation capabilities. Qubit rotations are implemented differently depending on the sets of qubits required, and on the axis of rotation chosen. For many applications rotating all qubits collectively is required. This can be efficiently implemented using a single, collective beam travelling axially along the spine of the trap, co-linearly with the ion string. The single-site-resolving addressing units described in section~\ref{subsec:Addressing} are used instead if only subsets of qubits need to be rotated.

The mechanism to drive the rotation itself is additionally different depending on the axis chosen. A rotation around an axis in the Bloch sphere's equatorial plane, that is between \Rx{} and \Ry{}, can be implemented via resonant excitation on the \SI{729}{\nano\meter} qubit transition. An optical phase imparted using the FAOMs is used to define the axis of rotation relative to the first pulse in a sequence, which may be defined arbitrarily. The remaining axis, \Rz{}, is instead driven by off-resonant AC-Stark pulses. Typically we perform $\pi/2$ rotations in \SI{15}{\micro\second}, limited by our current control electronics.

In order to characterize the fidelity of these single qubit rotations we perform randomized benchmarking on a single qubit~\cite{knill2008randomized}, assuming uncorrelated noise. A number of $n$ Clifford gates is executed, where each Clifford gate incurs an average cost of 1.875 \Rx{\pi}. At the end of the sequence the Clifford gate which inverts the sequence is added, thus creating an identity operation if not for gate errors. Here, we concatenate up to 100 Clifford gates and fit an exponential decay of the form $A\times p^n + 0.5$ to the population of the initial state. We assume that the population will decay to a value of 0.5 for a large number of gates. $A + 0.5$ is then the fidelity of state initialization. The error per Clifford gate is calculated as $R_\textrm{Clif} = (1-p)(1-1/d)$, with $d=2$ denoting the dimension of the Hilbert space for a single qubit.

First we characterize qubit rotations when using the global, axially-aligned beam that drives collective rotations. We obtain a Clifford gate fidelity of $F_\textrm{Clif} = \SI{99.83(1)}{\percent}$ or $F_\textrm{gate} =\SI{99.91(1)}{\percent}$ per single-qubit rotation. We perform the same measurement on a single ion with the tightly-focused beam from the microoptics addressing unit, and obtain $F_\textrm{Clif} = \SI{99.75(2)}{\percent}$ and $F_\textrm{gate} =\SI{99.86(1)}{\percent}$, respectively. 

\subsection{\MS{} gate and entanglement generation}

The ability to generate entanglement in a deterministic way is key for quantum computation~\cite{penrose1998quantum, jozsa2003role, datta2007role}, and the missing component to our complete gate set. Different types of entangling gates have been proposed and demonstrated for trapped-ion qubits~\cite{Cirac:1995, staanum2004geometric, kim2008geometric}. Here, we utilize the {\MS} gate~\cite{Sorensen:1999, sorensen2000entanglement} which generates entanglement through spin-dependent forces generated by a bichromatic light field slightly detuned from a vibrational mode. However, quantifying the degree and depth of entanglement in many-body quantum systems generated by any means remains a challenging task~\cite{amico2008entanglement}. Entanglement witnesses~\cite{bourennane2004experimental, sorensen2001entanglement} are often used, where the observable crossing a given threshold guarantees multi-partite entanglement of a given depth. Greenberger-Horne-Zeilinger (GHZ) states are a class of maximally-entangled Schr{\"o}dinger's cat states that have the fortunate features of being natively implemented by the \MS{} gate, and providing a particularly easy-to-measure entanglement witness. GHZ states further are highly susceptible to errors and decoherence~\cite{monz201114, omran2019generation, mooney:2021}, and as such provide a sensitive probe for characterizing the performance of our compound system.

\begin{figure}[h!]
    \centering
    \includegraphics[width=8.6cm]{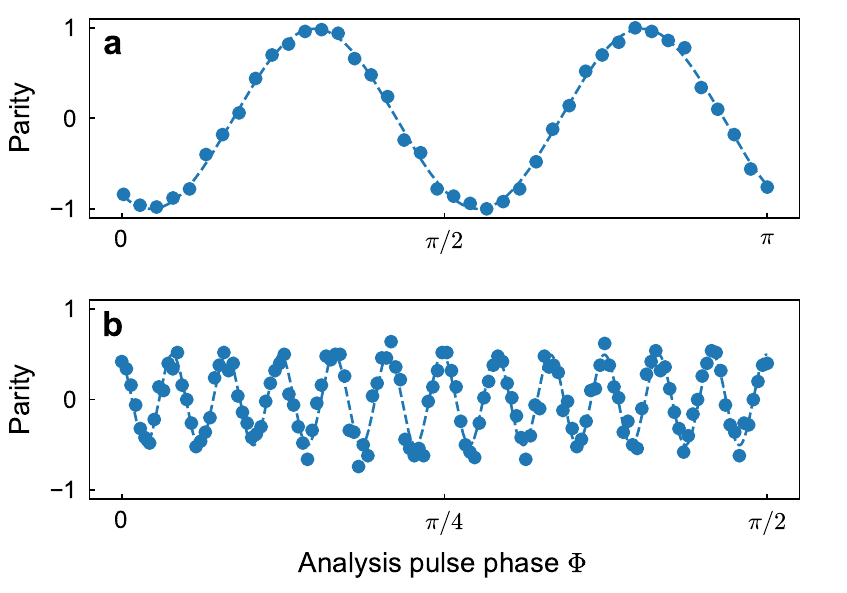}
    \caption{\textbf{a} Parity oscillations after a {\MS} gate on two ions followed by an evaluation pulse with phase $\Phi$. The amplitude of the parity oscillations is $C_2 = \SI{0.995(11)}{}$ with 100 measurements per data point. Together with populations $P_2(SS,DD) = \SI{0.9988(5)}{}$ we determine a fidelity of $F_2 = \SI{0.997(6)}{}$. \textbf{b} Parity oscillations of a 24-ion GHZ state with 100 measurements per data point. The parity contrast drops to $C_{24} = \SI{0.501(13)}{}$, the population to $P_{24}(S\ldots S,D\ldots D)= \SI{0.588(6)}{}$. We obtain a fidelity of $F_{24} = \SI{0.544(7)}{}$; more than 6 standard deviations above the 24-partite entanglement threshold of 0.5. Measurement uncertainties are below marker size.}
    \label{fig:MSGate}
\end{figure}

\begin{figure}[h!]
    \centering
    \includegraphics[width=8.6cm]{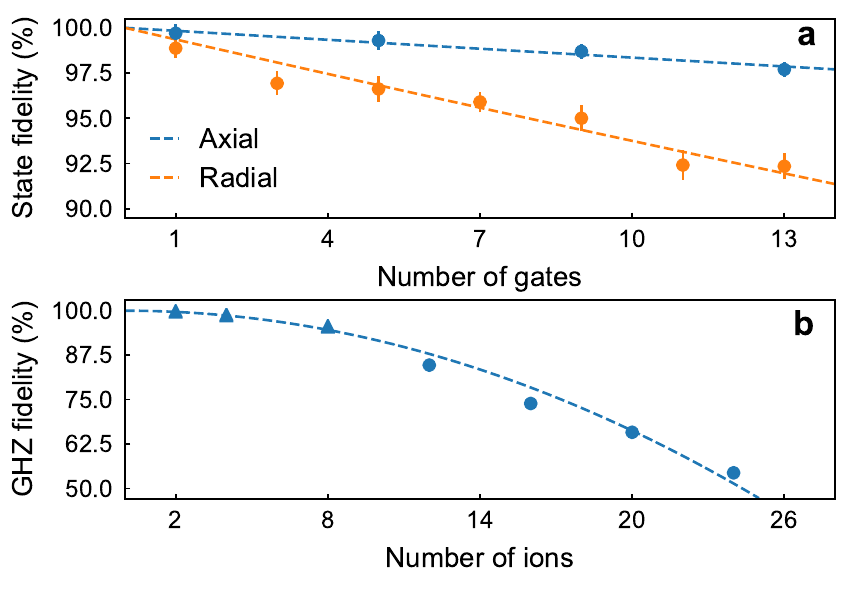}
    \caption{\MS{} gate performance. \textbf{a} Decay of overall state fidelity after repeated, odd-integer application of $\approx\SI{200}{\micro\second}$ \MS{} interactions for axial and radial two-ion gates. We infer a single-gate state fidelity of \SI{0.9983(1)}{} per axial gate and \SI{0.9936(3)}{} per radial gate. \textbf{b} Axial \MS{} gates on linear ion string for different ion numbers. Pairs of ions are successively added to the existing string and the measurement repeated. \mycircle{myblue} measurements are performed at an axial center-of-mass mode frequency of $\omega_\textrm{ax} = 2\pi\times\SI{234}{\kilo \hertz}$, to ensure the formation of a linear string for large number of ions, and were taken consecutively. \mytriangle{myblue} measurements are performed at an axial center-of-mass mode frequency of $\omega_\textrm{ax} = 2\pi\ \times\ $\SIrange{200}{1000}{\kilo \hertz} on different days. Measurement uncertainties are below marker size.
    }
    \label{fig:MSGate_nion_ngate}
\end{figure}

First, we implement \MS{} gates using a collective, axially-oriented beam on ion strings from 2 to 24 ions. The state fidelity $F$ after the entangling operation provides the entanglement witness with $F > 0.5$ guaranteeing full-depth entanglement. We characterize the operation using a two-ion crystal. The fidelity is directly inferable from the populations in the $\ket{S,S} = \ket{1, 1}$ and $\ket{D,D} = \ket{0, 0}$ states with \mbox{$P_2(SS,DD) = \SI{0.9988(5)}{}$} as shown in Fig.~\ref{fig:MSGate} \textbf{a}, as well as their coherences~\cite{monz201114}. The coherences are obtained from observing the oscillations in the state's parity as the phase of an analysis $\pi/2$ pulse is varied before projective measurement. Measuring the amplitude of this oscillation yields $C_2 = \SI{0.995(11)}{}$~\cite{leibfried2005creation, monz201114}. From those we calculate a single-gate state fidelity \mbox{$F_2 = (P_2 + C_2)/2$} with \mbox{$F = \SI{0.997(6)}{}$}. This fidelity includes all errors from state preparation and measurement (SPAM) and in particular is achieved without post-selection. 
A single two-ion \MS{} gate typically takes \SI{200}{\micro \second}. Sustaining the interaction for odd-integer multiples of this duration implements repeated \MS{} gates. Measuring the fidelity after different numbers of gates, and fitting an exponential decay as shown in Fig.~\ref{fig:MSGate_nion_ngate} \textbf{a} yields a simple estimator of the state fidelity with \SI{0.9983(1)}{} per gate.  

Larger multi-partite entangled states are subsequently produced by applying a single {\MS} gate to multiple ions. We demonstrate the generation of GHZ states up to 24 qubits. Measured fidelities are plotted in Fig. \ref{fig:MSGate_nion_ngate} \textbf{b}, where pairs of ions are successively added to the existing string and the measurement is repeated. For 24 ions we measure \mbox{$P_{24}(S\ldots S,D\ldots D)= \SI{0.588(6)}{}$}, \mbox{$C_{24} =\SI{0.501(13)}{}$} and \mbox{$F_{24} = \SI{0.544(7)}{}$}, which is again not corrected for SPAM errors and without post-selection, and is more than six standard deviations above the threshold for 24-partite entanglement~\cite{monz201114}. The parity oscillations are plotted in Fig.~\ref{fig:MSGate} \textbf{b}. To the best of our knowledge this is the largest GHZ state as well as the largest fully-entangled state that has been generated in any system without error mitigation or post-selection.

In the data shown above {\MS} gates were carried out on the axial modes of vibration. In order to create a fully programmable system, we use the radial modes of vibration for arbitrary-pair, addressed entangling operations on specific qubits in long ion strings. Characterizing the creation of states with addressed operations naturally leads to the question of fidelity metric; with two prominent candidates. The first is the fidelity of, say, GHZ state production where the non-addressed sub-register is ignored. This will quantify the action of an entangling gate, but is oblivious to what this operation does to the rest of the register. The second, more stringent choice, would be to determine the fidelity of performing an entangling operation on the addressed qubits without affecting the idling qubits. This is then the overlap of the full register state with the ideal register state, rather than sub-registers. Given that crosstalk is unavoidable we elect to chose the second metric.

Initial tests were carried out with the microoptics addressing unit, addressing the two outer ions of a three ion crystal. The {\MS} gate with the radially-oriented addressing beam yielded a register overlap fidelity of $F_2 = \SI{0.989(5)}{}$ for a single gate. Again concatenating multiple gates and fitting an exponential decay yields a fidelity of $F'_2 = \SI{0.9936(3)}{}$ per gate as plotted in Fig. \ref{fig:MSGate_nion_ngate} \textbf{a}. $F_2$ is the fidelity of {\MS} gate including SPAM errors, while $F'_2$ is approximately the pure fidelity per gate operation without SPAM errors.

\begin{figure}
    \centering
    \includegraphics[width = 6.02 cm]{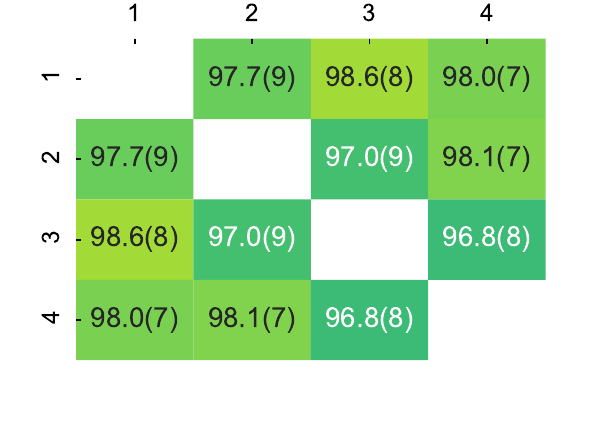}
    \caption{Full register overlap fidelity of states produced using AOD addressing unit in 4-ion crystal. There is no meaningful distinction between pair (1,4) and (4,1) as opposed to in single-site addressing. Consequently, the matrix is symmetric.}
    \label{fig:AOD_MS_Pairs}
\end{figure}

With the AOD addressing unit we measure all pairwise-entangling gates in a 4-ion crystal. We calibrate the gate on the two outer ions and use the same set of parameters for all gates. The register overlap fidelities are shown in Fig.~\ref{fig:AOD_MS_Pairs}. We achieve fidelities in the range from 0.969(7) to 0.986(8). Cumulative population in the nominally non-addressed sub-register for these measurements was below \SI{0.2}{\percent}.

We anticipate that further improvements on radial mode stability, cooling of radial modes, addressing unit and mode disentanglement~\cite{milne2020phase, shapira2018robust, blumel2021efficient}, should increase the fidelity of the addressed gates.

\section{Conclusion}
\label{sec:Conclusion}

In this manuscript we have provided a detailed description of the experimental implementation of a compact, trapped-ion quantum computing demonstrator situated in two 19-inch racks. We presented mechanical, optical and electrical systems along with characterizing experiments. This experimental platform improves upon conventional hardware implementations in terms of modularity, integration, and remote control. In our characterization measurements we find the system performance to be on par with conventional, laboratory-based hardware implementations in terms of experimentally relevant performance criteria. We find that mechanical stability, optical readout and addressing performance, heating rates, coherence times, and {\MS} entangling fidelities in the current implementation do not suffer relative to traditional optical table setups. Using the compound system we are able to produce maximally-entangled Greenberger–Horne–Zeilinger states with up to 24 qubits with a fidelity of \SI{54.4(7)}{\percent}. To our knowledge this is the largest maximally-entangled state yet generated in any system without the use of error mitigation or post-selection, and demonstrates the capabilities of our system. 

In addition, we presented site-selective qubit operations to implement a complete gate set for universal quantum computation using two distinct approaches to addressing: A microoptics approach with fixed, rigid waveguides, and an acousto-optic deflector approach. Both of these approaches offer advantages over the other in particular settings.

The microoptics approach readily offers itself for simultaneous multi-site addressing without producing off-axis spots which can lead to resonant and off-resonant crosstalk. The power scaling in such a scenario is linear in the number of channels, assuming a suitable power distribution system is at hand. Parallel radial sideband cooling is one direct beneficiary of such capabilities, as is direct generation of interaction between more than two qubits. Individual control over amplitude, phase, and frequency of each channel is a fundamental advantage of this approach but requires one FAOM per qubit. The positional stability is not affected by the oscillator stability of RF sources, and extension to larger registers are not limited by the speed of sound or similar quantities such as in AOD-based devices.

The AOD approach on the other hand is technologically simpler, thus offering superior performance at this stage. Optical quality of macroscopic components is often also superior to microoptics, in particular in prototyping scenarios such as here, which reduces abberations. The addressing is inherently flexible, such that it can be adjusted for qubit registers from 2 ions to 40 ions in our configuration. Adjustment of power and optical phase of individual channels is possible without an optical modulator per ion, which significantly reduces the technological overhead compared to the microoptics approach. This unit is fed by a single fiber, and no prior power distribution capabilities are required. The switching speed in AODs is limited ultimately by the speed of sound in the deflecting crystal, which therefore also limits the speed at which operations can be performed. The quadratic power loss for multi-site addressing, and off-axis spots limit this technology to the simultaneous addressing of a small number of ions. 

With both of these approaches we demonstrate single-qubit and pairwise-entangling gates on registers up to 10 ions. From randomized benchmarking we obtain a fidelity of $F_\textrm{gate} =\SI{99.86(1)}{\percent}$ per addressed single-ion gate. Measurements of resonant crosstalk were shown to be below 1\% across the entire 10-ion register with the AOD approach, while non-resonant crosstalk was measured to be less than $1.25\times10^{-4}$ in the same string. Together with the pairwise entangling operations with fidelities between 97\% and 99\% we show all the basic operations for a fully programmable quantum system. Benchmarking of larger registers, and with more complete suites of benchmarking tools~\cite{wright2019benchmarking} will be undertaken as part of the next round of hardware integration and software improvements.

These near and mid-term upgrades to the hardware and software stacks will further improve upon the demonstrator's capabilities. Use of an external master oscillator to generate the narrow-linewidth qubit laser will no longer be required after installation of the, compact diode-laser source which is currently under construction as part of the AQTION collaboration. Similarly, single-site addressing capabilities will be improved in mode quality and number of channels. This will allow the setup to implement more complex algorithms by moving from axial gates to radial quantum gates enhanced by established quantum control techniques~\cite{werschnik2007quantum, khodjasteh2009dynamically}. Upgrades to M-ACTION, as well as complimentary developments to the control and remote access software stack will enable easier access to the demonstrator capabilities in a hardware-agnostic fashion.

\begin{figure}[h!]
    \centering
    \includegraphics[width = 8.6cm]{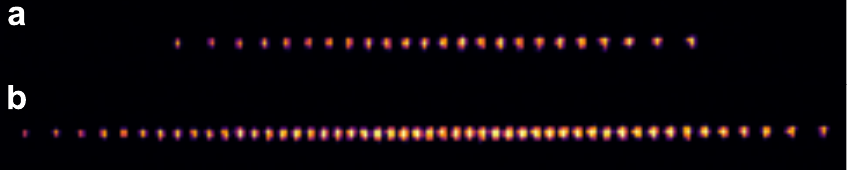}
    \caption{Ion images of \textbf{a} 24 and \textbf{b} 50 ions in the demonstrator. The 24-ion chain was used to demonstrate 24-partite entanglement within the setup without the use of post-selection or error mitigation. 50-ion chains are the mid-term control target and can already be trapped and cooled. Non-uniform brightness stems from the finite size of the detection beam.}
    \label{fig:outlook}
\end{figure}

Already, the device presented is capable of operating with qubit numbers on par with state-of-the-art conventional laboratory setups. An image of such a qubit register is shown in Fig.~\ref{fig:outlook} \textbf{a}. The mid-term upgrades should enable us to increase this number to the AQTION control target of 50 qubits and beyond. We have already demonstrated basic capabilities of control in larger qubit chains as shown in Fig.~\ref{fig:outlook} \textbf{b}, with a 50 ion chain already crystallized in our trap. In the long term, we hope that the demonstrator's features and engineering solutions mean that ion trap-based quantum computers may be situated in any space with reasonable environmental stability and vibrational level. Quantum computation with qubit count exceeding 100 is feasible based on our architecture and this characterization of its first implementation.

\section{Acknowledgements}
We gratefully acknowledge funding from the EU H2020-FETFLAG-2018-03 under Grant Agreement no. 820495. We also acknowledge support by the Austrian Science Fund (FWF), through the SFB BeyondC (FWF Project No.\ F7109), and the IQI GmbH.  This project has received funding from the European Union’s Horizon 2020 research and innovation programme under the Marie Skłodowska-Curie grant agreement No 840450. P.S. and M.M. acknowledge support from the Austrian Research Promotion Agency (FFG) contract 872766.  P.S., T.M. and R.B. acknowledge funding by the Office of the Director of National Intelligence (ODNI), Intelligence Advanced Research Projects Activity (IARPA), via US ARO grant no. W911NF-16-1-0070 and W911NF-20-1-0007, and the US Air Force Office of Scientiﬁc Research (AFOSR) via IOE Grant No. FA9550-19-1- 7044 LASCEM.

All statements of fact, opinions or conclusions contained herein are those of the authors and should not be construed as representing the official views or policies of the funding agencies.

\section*{References}

%

\end{document}